\documentclass[submission, Proceedings]{SciPostSub}
\hypersetup{
    colorlinks,
    linkcolor={red!50!black},
    citecolor={blue!50!black},
    urlcolor={blue!80!black}
}

\usepackage[bitstream-charter]{mathdesign}
\DeclareSymbolFont{usualmathcal}{OMS}{cmsy}{m}{n}
\DeclareSymbolFontAlphabet{\mathcal}{usualmathcal}

\usepackage{nicefrac}
\usepackage[multiple]{footmisc}

\usepackage{footnote}

\usepackage{mhchem}

\usepackage[bitstream-charter]{mathdesign}
\usepackage{mhchem}
\usepackage{subcaption}
\usepackage{slashed}

\makesavenoteenv{tabular}
\makesavenoteenv{table}
\DeclareSymbolFont{usualmathcal}{OMS}{cmsy}{m}{n}
\DeclareSymbolFontAlphabet{\mathcal}{usualmathcal}
\renewcommand{\eqref}[1]{(\ref{#1})}

\newcommand{\tabref}[1]{Table~\ref{#1}}
\newcommand{\secref}[1]{Section~\ref{#1}}
\newcommand{\MeV}{\,\mbox{MeV}}

\newcommand{\cL}{{\cal L}}
\renewcommand{\Re}{{\rm Re}}
\renewcommand{\Im}{{\rm Im}}
\newcommand{\chpt}{$\chi$PT}
\numberwithin{equation}{section}
\numberwithin{figure}{section}
\numberwithin{table}{section}

\definecolor{whcol}{rgb}{0.368417,0.506779,0.709798}
\definecolor{wcol}{rgb}{0.1,0.2,0.95}
\definecolor{xxcol}{rgb}{0.880722,0.611041,0.142051}
\definecolor{emhcol}{rgb}{0.560181,0.691569,0.194885}
\definecolor{emcol}{rgb}{0.1,0.6,0.1}
\definecolor{bsmcol}{rgb}{0.922526,0.385626,0.209179}
\definecolor{scol}{rgb}{0.528488,0.2,0.701351}

\begin{document}

\begin{center}{\Large \textbf{\color{scipostdeepblue}{
%%%%%%%%%% TODO: TITLE Paste title here
% multiline titles: end with a \\ to regularize line spacing
A theory vade mecum for PSI experiments\\
}}}\end{center}

\begin{center}
\textbf{
G. Colangelo\textsuperscript{1},
F. Hagelstein\textsuperscript{2}, 
A. Signer\textsuperscript{2,3$\star$} and
P. Stoffer\textsuperscript{4}
}
\end{center}

\begin{center}
{\bf 1} Albert Einstein Center for Fundamental Physics, Institute for
Theoretical Physics, University of Bern,
Switzerland
\\
{\bf 2} Paul Scherrer Institut, 5232 Villigen PSI, Switzerland
\\
{\bf 3} University of Zurich, Physik-Institut, 8057 Zurich, Switzerland
\\
{\bf 4} University of Vienna, Faculty of Physics, Boltzmanngasse 5, 1090
Vienna, Austria 
%%%%%%%%%% END TODO: AFFILIATIONS
%%%%%%%%%% TODO: EMAIL (OPTION) provide email address of corresponding author
\\[\baselineskip]
$\star$ \href{mailto:adrian.signer@psi.ch}{\small \sf adrian.signer@psi.ch}
%%%%%%%%%% END TODO: EMAIL (OPTION)
\end{center}

\definecolor{palegray}{gray}{0.95}
\begin{center}
\colorbox{palegray}{
 \begin{tabular}{rr}
 \begin{minipage}{0.05\textwidth}
   \includegraphics[width=24mm]{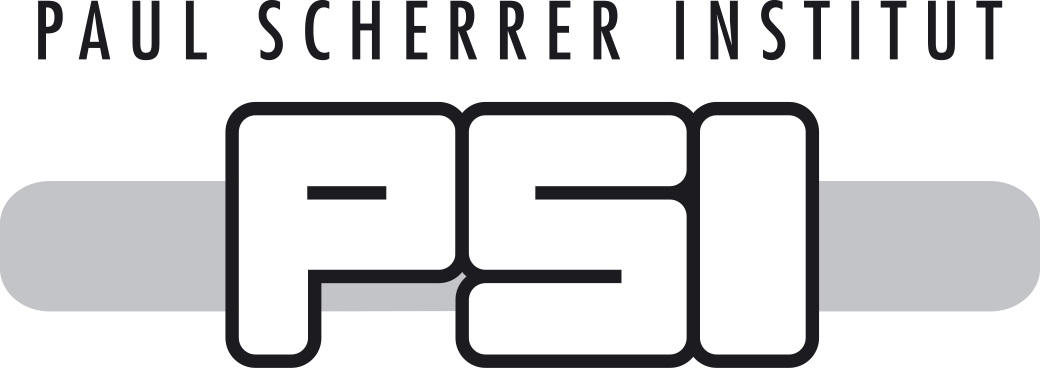}
 \end{minipage}
 &
 \begin{minipage}{0.82\textwidth}
   \begin{center}
   {\it Review of Particle Physics at PSI}\\
   \doi{10.21468/SciPostPhysProc.5}\\
   \end{center}
 \end{minipage}
\end{tabular}
}
\end{center}

\section*{\color{scipostdeepblue}{Abstract}}
{\bf
%%%%%%%%%% TODO: ABSTRACT Paste abstract here
This article gives a compact introduction and overview of the theory underlying the experiments described in the rest of this review.
}

\begin{center}
\begin{tabular}{lr}
\begin{minipage}{0.56\textwidth}
\raisebox{-1mm}[0pt][0pt]{\includegraphics[width=12mm]{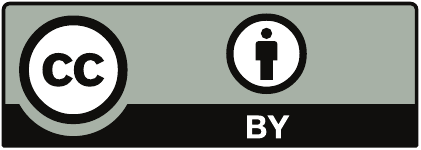}}
%%%%%%%%%% TODO: COPYRIGHT Include the first author's initials and last name
{\small Copyright G. Colangelo {\it et al}. \newline
%%%%%%%%%% END TODO: COPYRIGHT
This work is licensed under the Creative Commons \newline
\href{http://creativecommons.org/licenses/by/4.0/}{Attribution 4.0 International License}. \newline
Published by the SciPost Foundation.
}
\end{minipage}
&
\begin{minipage}{0.44\textwidth}
\noindent\begin{minipage}{0.68\textwidth}
%%%%%%%%%% TODO: DATES
{\small Received 17-05-2021 \newline Accepted 21-07-2021 \newline Published 06-09-2021
%%%%%%%%%% END TODO: DATES
%%%%%%%%%% TODO: DOI
}
\end{minipage}
\begin{minipage}{0.25\textwidth}
\begin{center}
\href{https://crossmark.crossref.org/dialog/?doi=10.21468/SciPostPhysProc.5.005&amp;domain=pdf&amp;date_stamp=2021-09-06}{\includegraphics[width=7mm]{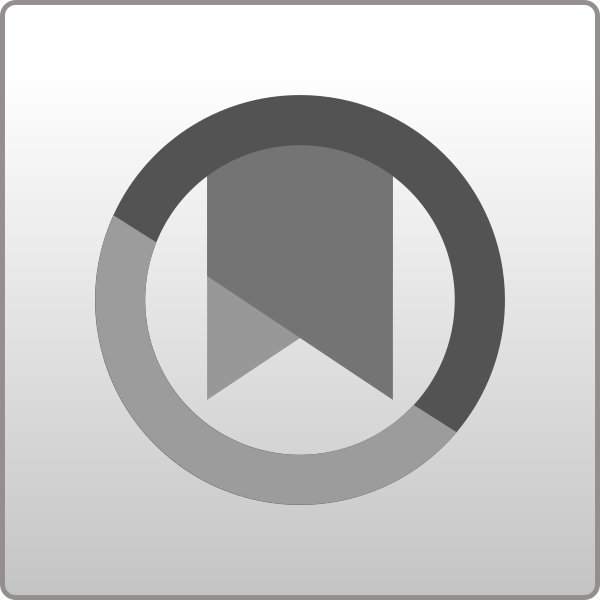}}\\
\tiny{Check for}\\
\tiny{updates}
\end{center}
\end{minipage}
\\\\
\small{\doi{10.21468/SciPostPhysProc.5.005}
%%%%%%%%%% END TODO: DOI
}
\end{minipage}
\end{tabular}
\end{center}

%%%%%%%%%% TODO: LINENO Activate linenumbers during proofs stage
%\linenumbers

%%%%%%%%%% TODO: TOC include a table of contents (optional)
\vspace{10pt}
\noindent\rule{\textwidth}{1pt}
%\tableofcontents
%\noindent\rule{\textwidth}{1pt}
\vspace{10pt}

%%%%%%%%% TODO: CONTENTS Contents come here, starting from first \section
\setcounter{section}{5}
\label{th:sec}
\subsection{Introduction}\label{th:intro}

The purpose of this article is to give a broad overview of the theory
background to the experi\-ments that have been and are carried out at
the Paul Scherrer Institute. Space limitations make it impossible to
go into depth or provide a self-contained theoretical summary. Much
more modestly, we aim to put the experiments into context and provide
key references for further reading. The experiments we refer to are
listed in \tabref{sectiontable} and they will be described in greater
detail in separate sections/articles of the Review of Particle Physics
at PSI~\cite{section06, section16, section07, section08, section19,
  section20, section29, section09, section21, section22, section23,
  section17, section18, section14, section26, section15, section27,
  section28, section10, section11, section12, section24,
  section25}. These experiments either
lead to precise determinations of physical parameters required as
input for other experiments (e.g., muon life time, pion mass), or search for physics beyond the Standard
Model (BSM). The BSM searches proceed along different frontiers. One way to search for new physics is to consider physical observables whose Standard
Model (SM) contributions either vanish or are too small to be experimentally
accessible. In other words, they are identical to zero for practical purposes. Examples are 
charged lepton-flavor violating (cLFV) muon decays or a permanent neutron
electric dipole moment (EDM). To put constraints on the branching ratios of BSM decays, one has to observe a large number of decays. This is, thus, called a search at the intensity frontier. Another way to search for new
physics is to consider precision observables and search for deviations from the SM expectations. Prominent examples are the precision QED tests with muonium, as well as the precision laser spectroscopy experiments with muonic atoms. These are, thus, called searches at the precision frontier. The low-energy experiments at PSI are complementary to the experiments at LHC, which sit at the energy frontier.

%In the following we aim to give meaning to these three statements.

After a general overview of the theoretical methods applied to
describe the processes and bound states in \tabref{sectiontable}, we
will, in turn, consider the muon, the proton, nucleons and nuclei, the
free neutron, and the pions.

\setlength{\tabcolsep}{0.5em}
{\renewcommand{\arraystretch}{1.4}
\begin{table}
  \caption{\label{sectiontable} Processes and particles (bound states) that are investigated at PSI, where the driving interaction to be studied is indicated by the color as follows: \textcolor{bsmcol}{BSM},
    \textcolor{wcol}{weak},
    \textcolor{whcol}{weak and try to learn about strong},
    \textcolor{emcol}{EM},
    \textcolor{emhcol}{EM and try to learn about strong},
    \textcolor{scol}{strong}. In addition the mass or \textcolor{gray}{charge
      radius} of particles are measured. The section number refers to the Review of Particle Physics at PSI.} 
 \begin{tabular}{rl|r|l}
      & experiment & section & process / particles / (bound states) \\[3pt]
    \hline
          \hypertarget{ht:section06}{\cite{section06}} &
    muon decay & 6  &
    \textcolor{wcol}{$\mu^+\to e^+ \nu_e\bar{\nu}_\mu$} \\
    \hypertarget{ht:section16}{\cite{section16}} &
    MuLan & 16 &
    \textcolor{wcol}{$\mu^+\to e^+ \nu_e\bar{\nu}_\mu$} \\
    \hypertarget{ht:section07}{\cite{section07}} &
    SINDRUM &  7 &
    \textcolor{bsmcol}{$\mu^+\to e^+\, ee$},
    \textcolor{wcol}{$\mu^+\to e^+ \nu_e\bar{\nu}_\mu\, ee$},
     \textcolor{wcol}{$\pi^+\to e^+ \nu_e \, ee$},
     \textcolor{emhcol}{$\pi^0\to e e$} \\
    \hypertarget{ht:section08}{\cite{section08}} &
    SINDRUM~II & 8 &
    \textcolor{bsmcol}{$\mu^-\, \ce{_Z^A}N \to e^-\, \ce{_Z^A}N$} \quad
    for  Au, Pb, Ti\\ 
    \hypertarget{ht:section19}{\cite{section19}} &
    MEG &  19 &
    \textcolor{bsmcol}{$\mu^+\to e^+ \gamma$},
    \textcolor{wcol}{$\mu^+\to e^+ \nu_e\bar{\nu}_\mu\gamma $},
    \textcolor{bsmcol}{$\mu^+\to e^+ X \to  e^+\gamma\gamma $} \\
    \hypertarget{ht:section20}{\cite{section20}} &
    Mu3e & 20 &
    \textcolor{bsmcol}{$\mu^+\to e^+\, ee$},
    \textcolor{wcol}{$\mu^+\to e^+ \nu_e\bar{\nu}_\mu\, ee$} \\
    \hypertarget{ht:section29}{\cite{section29}} &
    Mspec, Mu-Mass & 29 &
    \textcolor{emcol}{$M=(\mu^+ e^-)$}, $\mu^+$\\ 
    \hypertarget{ht:section09}{\cite{section09}} &
    MACS & 9 &
    \textcolor{bsmcol}{$M=(\mu^+ e^-) \leftrightarrow
    \bar{M}=(\mu^- e^+)$} \\
    \hline
    \hypertarget{ht:section21}{\cite{section21}} &
    CREMA  & 21 &
    \textcolor{emhcol}{$(\mu^- p)$},
    \textcolor{emhcol}{$(\mu^- d)$},
    \textcolor{emhcol}{$(\mu^- \mbox{He})$},
     \quad \textcolor{gray}{$p$},
      \textcolor{gray}{$d$}, \textcolor{gray}{He}  \\
    \hypertarget{ht:section22}{\cite{section22}} &
    muX  & 22 &
    \textcolor{emhcol}{$(\mu^{-}\, \ce{_Z^A}N)$},
    \quad \textcolor{gray}{$\ce{^{248}_{96}}$Cm},
    \textcolor{gray}{$\ce{^{226}_{88}}$Ra} \\
    \hypertarget{ht:section23}{\cite{section23}} &
    MUSE & 23 &
    \textcolor{emhcol}{$e^\pm p \to e^\pm p$},
    \textcolor{emhcol}{$\mu^\pm p \to \mu^\pm p$} \\
    \hypertarget{ht:section17}{\cite{section17}} &
    MuCap  & 17 &
    \textcolor{whcol}{$\mu^- p \to \nu_\mu n$} \\
    \hypertarget{ht:section18}{\cite{section18}} &
    MuSun  & 18 &
    \textcolor{whcol}{$\mu^- d \to \nu_\mu n n$} \\
     \hypertarget{ht:section14}{\cite{section14}} &
     pionic hydrogen & 14 &
    \textcolor{emhcol}{$(\pi^- p)$},  \textcolor{emhcol}{$(\pi^- d)$}\\   
    \hypertarget{ht:section26}{\cite{section26}} &
    pionic helium & 26 &
    \textcolor{emcol}{$(\pi^- e^-\ ^4\mbox{He}^{++})$}, $\pi^-$\\
    \hline
    \hypertarget{ht:section15}{\cite{section15}} &
    nTRV & 15 &
    \textcolor{wcol}{$n\to p e^- \bar\nu_e$} \\
    \hypertarget{ht:section27}{\cite{section27}} &
    nEDM  & 27 &
    \textcolor{scol}{$n$},
    \textcolor{bsmcol}{$n$} \\
    \hypertarget{ht:section28}{\cite{section28}} &
    indirect nEDM  & 28 &  \textcolor{bsmcol}{$n$ / dark matter / exotic} \\
    \hline
   \hypertarget{ht:section10}{\cite{section10}} &
    negative pions & 10 &
    \textcolor{emcol}{$(\pi^- p)$}, $\pi^-$ \\
    \hypertarget{ht:section11}{\cite{section11}} &
    positive pions  & 11 &
    \textcolor{wcol}{$\pi^+\to \mu^+ \nu_\mu$}, $\pi^+$, $\nu_\mu$\\ 
    \hypertarget{ht:section12}{\cite{section12}} &
    neutral pions & 12 &
    \textcolor{wcol}{$\pi^- p \to \pi^0 n$},  $\pi^0$ \\
    \hypertarget{ht:section24}{\cite{section24}} &
    PiBeta & 24 &
    \textcolor{wcol}{$\pi^+\to \pi^0 e^+ \nu_e$},
    \textcolor{wcol}{$\pi^+\to e^+ \nu_e\, (+ \gamma)$},
%    \textcolor{wcol}{$\pi^+\to e^+ \nu_e \gamma$},
    \textcolor{wcol}{$\mu^+\to e^+ \nu_e\bar{\nu}_\mu\gamma $} 
    \\
    \hypertarget{ht:section25}{~\cite{section25}} &
    PEN & 25 &
    \textcolor{wcol}{$\pi^+\to e^+ \nu_e\, (+ \gamma)$},
%    \textcolor{wcol}{$\pi^+\to e^+ \nu_e \gamma$},
    \textcolor{wcol}{$\mu^+\to e^+ \nu_e\bar{\nu}_\mu\gamma $}  \\
    \hline
  \end{tabular}
\end{table}
}

\subsection{Overview}\label{th:secoverview}

The experiments we are primarily concerned with involve low-energy interactions of
electrons, muons, protons, neutrons, and pions. In Section~\ref{th:secoverviewSM} we
first describe these interactions in the SM before we discuss the generalization to BSM
scenarios in Section~\ref{th:secoverviewBSM}. While the theoretical methods for these
cases are
dominated by perturbative expansions in the couplings, Section~\ref{th:secoverviewHAD} is
devoted to hadronic effects that often play an important part in low-energy experiments.

\subsubsection{Standard Model at low energies}\label{th:secoverviewSM}

In the SM the dynamics of the particles listed above is described by the gauge theory of strong and electroweak interactions. In view of the large masses of the Higgs and weak gauge bosons, the weak part of the SM Lagrangian is essentially frozen at low energies (it will later be considered as a small correction). In this regime, the SM reduces to the standard QED and QCD Lagrangian
\begin{align}
  \label{th:LagQEDQCD}
  \cL_\text{QED+QCD} = \sum_f \bar{f} \left(i\slashed{D} - m_f\right) f
  -\frac{1}{4} F_{\alpha\beta} F^{\alpha\beta}
  -\frac{1}{4} G_{\alpha\beta} G^{\alpha\beta} 
%+ \frac{g^2 \theta}{32\pi^2} \tilde{G}_{\alpha\beta} G^{\alpha\beta}
  \, ,
\end{align}
where the electromagnetic and gluonic field-strength tensors are expressed in terms
of the photon and gluon fields, $A^\alpha$ and $G^\alpha$, as $F^{\alpha\beta}= \partial^\alpha
A^\beta - \partial^\beta A^\alpha$, $G^{\alpha\beta} = \partial^\alpha
G^\beta - \partial^\beta G^\alpha - i g_s [G^\alpha, G^\beta]$, and where for clarity we have omitted gauge-fixing and ghost terms. The sum runs over all fermions of
mass $m_f$, electric charge $e\,Q_f$, and color charge $g_s t^a_f$, and the covariant derivative acts on
the fermion fields as $D_\alpha f = (\partial_\alpha - i e Q_f
A_\alpha - i g_s t_f^a G_\alpha^a) f$. For $f=\ell\in\{e,\mu,\tau\}$ we have $Q_\ell = -1$ and $t_\ell^a = 0$,
whereas for quarks $Q_u=2/3$, $Q_d=-1/3$, and $t_{u,d}^a = \lambda^a/2$ with Gell-Mann matrices $\lambda^a$. In several experiments of interest here the photon acts as a probe: it is coupled to the
electromagnetic current $J_\text{em}^\alpha$ as
\begin{align}
  \label{th:JQED}
  \cL_\text{QED}^\text{int} = e\, A_\alpha J_\text{em}^\alpha
  \equiv  e\, A_\alpha \sum_f Q_f \bar{f}\gamma^\alpha f \,.
\end{align}
If we use~\eqref{th:LagQEDQCD} to compute the matrix element of
$J_\text{em}^\alpha$ between two states of pointlike leptons $\ell$
with momenta $p_1$ and $p_2=p_1+q$, we find
\begin{align}
  \label{th:leptonJem}
  \langle \ell(p_2)| J_\text{em}^\alpha |\ell(p_1)\rangle
  = \bar{u}(p_2,m_\ell)\left(
  F_1^{(\ell)}(q^2)\, \gamma^\alpha
  + F_2^{(\ell)}(q^2) \frac{i\,\sigma^{\alpha\beta}  q_\beta}{2\,m_\ell} \right)
  u(p_1,m_\ell) \, ,
\end{align}
where $u$ and $\bar{u}$ are the usual spinors. The decomposition~\eqref{th:leptonJem} directly follows from the Lorentz and $U(1)_\text{em}$ gauge symmetries of the theory and is valid beyond perturbation theory.
While $F^{(\ell)}_1$ is related to the electric charge, $F^{(\ell)}_2$ is
related to the anomalous magnetic moment (AMM) of $\ell$ as
\begin{align}
  \label{th:F2al}
  F_2^{(\ell)}(0) = a_\ell = \frac{(g-2)_\ell}{2}\, .
\end{align}
In contrast to the leptons, quarks do not appear as free particles in nature, but are confined inside hadrons by the strong interaction. The general principles on which the decomposition~\eqref{th:leptonJem} is based, also hold for non-pointlike particles, such as the nucleons $N\in\{p,n\}$
\begin{align}
  \label{th:nucleonJem}
  \langle N(p_2)| J_\text{em}^\alpha |N(p_1)\rangle
  = \bar{u}(p_2,m_N)\left(
  F_1^{(N)}(Q^2)\, \gamma^\alpha
  + F_2^{(N)}(Q^2) \frac{i\,\sigma^{\alpha\beta}  q_\beta}{2\,m_N} \right)
  u(p_1,m_N) \, ,
\end{align}
where we have introduced the common definition $Q^2\equiv -q^2$. A relation between the AMM and $F_2^{(N)}$ analogous to~\eqref{th:F2al} still holds. However, this quantity depends on strong dynamics, which at low energies cannot be computed in perturbation theory. 

In the case of the nucleons, often the electric and magnetic form
factors
\begin{align}
  \label{th:FtoG}
  G_E^{(N)}(Q^2)&\equiv F_1^{(N)}(Q^2) - \frac{Q^2}{4 m^2_N} F_2^{(N)}(Q^2), &
  G_M^{(N)}(Q^2)&\equiv F_1^{(N)}(Q^2) +  F_2^{(N)}(Q^2) , 
\end{align}
are used. In the limit of small $Q^2$ all form factors $F_i(Q^2)$ can be
understood as the Fourier transform of an extended classical `charge'
distribution $\rho_i(r)$ in the Breit frame where $q^\mu = (0,\vec{q})$. Upon expansion in small $Q^2$ we get
\begin{align}
  \label{th:chargedist}
   F_i(Q^2) = \int\!d^3\vec{r}\,  e^{- i\,\vec{q}\cdot\vec{r}}\,  \rho_i(r)
  = \int\!d^3\vec{r}\ \rho_i(r)
  - \frac{1}{6}  Q^2\, \int\!d^3\vec{r}\ r^2\,\rho_i(r)  + \ldots
\end{align}
This leads to a general expression for the second moment of the
charge distribution $\rho_i$
\begin{align}
  \label{th:radius}
  r_i^2 \equiv
  \frac{1}{N} \int\!d^3\vec{r}\ r^2\,\rho_i(r)
  = -6 \frac{1}{N} \frac{dF_i(Q^2)}{d Q^2}\bigg|_{Q^2=0} \, , \quad N = \left\{ \begin{array}{ll}
     1  & \text{ if } F_i(0) = 0 \, ,  \\
     F_i(0)  & \text{ else.}
  \end{array}\right.
\end{align}
The relation above is used for example to determine the
root-mean-square, $R_i=\sqrt{r_i^2}$, charge and magnetic radii of the
proton as well as the axial radius of the nucleon.

If we now consider the weak interactions, we must arrange fermions into
left-handed doublets and right-handed singlets. An important role for
low-energy processes is played by the charged weak current
\begin{align}
\label{th:Jcc}
J_\text{cc}^\alpha &= 
 \sum_\ell \bar{\nu}_\ell \gamma^\alpha P_L \ell 
+ \sum_{ij} V_{ij}\, \bar{u}_i \gamma^\alpha P_L {d_j},
\end{align}
which couples only to left-handed fermions, $P_L\equiv(1-\gamma_5)/2$.
In the sum over the quark-field terms, the CKM matrix $V_{ij}$
describes the flavor-changing effects of the weak interactions.
Including for completeness also the neutral weak current
$J_\text{nc}^\alpha$, the interactions of \eqref{th:JQED} are modified
to
\begin{align}
  \label{th:JEW}
  \cL_\text{EW}^\text{int} =  e\, A_\alpha J_\text{em}^\alpha 
 + \frac{g}{\sqrt{2}} \big(
     W_\alpha^+ J_\text{cc}^\alpha + \mbox{h.c.} \big)
 + g_Z\, Z_\alpha J_\text{nc}^\alpha \, ,
\end{align}
where $g=e/\sin\theta_W$, $g_Z = g/\cos\theta_W$ are the weak
$SU(2)_L$ couplings that can be expressed in terms of $e$ and the
electroweak mixing (Weinberg) angle $\theta_W$. At the typical energy
of processes considered here, much smaller than $m_W$ and $m_Z$, the
$W$ and $Z$ boson masses, we can integrate out the $W$ and $Z$ bosons
and adopt an effective field theory (EFT) approach. This results in
the Fermi theory of current-current interactions
\begin{align}
\label{th:L4F}
\cL_{4F} &= - \frac{4 G_F}{\sqrt{2}} \Big( J_\text{cc}^\alpha
  (J_\text{cc})_\alpha^\dagger + J_\text{nc}^\alpha
  (J_\text{nc})_\alpha \Big) \, ,
\end{align}
where $4\, G_F/\sqrt{2} = g^2/(2 m_W^2)$ is the matching (Wilson)
coefficient at tree level. Using \eqref{th:Jcc} (and the corresponding
expression for $J_\text{nc}^\alpha$) to express $\cL_{4F}$ in terms of
fermion fields we end up with vector contact interactions. They
correspond to dimension-6 four-fermion vector operators of the generic
form
\begin{align}
\label{th:opV4f}
 \big[O^{V,XY}_{\{\ell/q\}}\big]_{ijkl} &= \left( \bar\psi_i
 \gamma^\alpha P_X \psi_j \right) \left( \bar\psi_k \gamma_\alpha
 P_Y \psi_l \right)\, ,
\end{align}
where $X,Y\in\{L,R\}$ and $\{i,j,k,l\}$ are generation indices. The
notion `vector' refers to the Lorentz structure of the bilinears,
which in turn is closely related to the nature of the exchange
particle that is integrated out. Since the fermion fields $\psi_{i}$
can be quarks or leptons of any generation, there are in principle
quite a lot of different operators. However, only a subset of those
are generated by integrating out the $W$ and $Z$ fields. In
particular, there are no charged cLFV
operators due to an accidental symmetry of the SM.

Because the masses of the top quark and the Higgs boson are of the same
order as $m_W$, these fields can also be integrated out. Operators
beyond the four-fermion vector operators appear in the SM with an
additional suppression, such as scalar dimension-6 four-fermion
operators
\begin{align}
\label{th:opS4f}
 \big[O^{S,XY}_{\{\ell/q\}}\big]_{ijkl} &= \left( \bar\psi_i
 P_X \psi_j \right) \left( \bar\psi_k P_Y \psi_l \right)\, , \quad X, Y \in \{L,R\} \, ,
\end{align}
which are parametrically suppressed by Yukawa
couplings~\cite{Jenkins:2017jig}, or dimension-5 dipole operators (and
their Hermitian conjugate)
\begin{align}
\label{th:opD}
 \big[O^{D}_{\{\ell/q\}\gamma}\big]_{ij} &= \left( \bar\psi_i
 \sigma_{\alpha\beta} P_R \psi_j \right) F^{\alpha\beta}\, , \quad \big[O^{D}_{qG}\big]_{ij} = \left( \bar\psi_i
 \sigma_{\alpha\beta} G^{\alpha\beta} P_R \psi_j \right) \, ,
\end{align}
which appear at the loop level.  Thus, we arrive at an EFT that
consistently describes low-energy processes. It only contains fields
with masses much lower than $m_W$. In particular, the photon and the
gluons are the only gauge bosons present. The gauge symmetry of the
SM, $SU(3)_c\times SU(2)_L \times U(1)_Y$, is reduced to the gauge
symmetry of QCD and QED, $SU(3)_c\times U(1)_\text{em}$. The effect of
the heavy degrees of freedom of the SM is encoded in the Wilson
coefficients that multiply the operators, with $G_F$ in \eqref{th:L4F}
being one such example.

\subsubsection{Low-energy physics beyond the Standard Model}\label{th:secoverviewBSM}

%The SM described above is remarkably successful. However, in some cases the SM fails to %describe the physics that we observe. There is for instance evidence of BSM physics %coming  from astrophysical observations of Dark Matter (DM) and Dark Energy. 

Many of the experiments listed in Table~\ref{sectiontable} are motivated by the search for new physics. One can think of a plethora of BSM scenarios. They rely on different interaction mechanisms, and can be roughly classified based on the masses of the BSM particles and their coupling strengths. 

Light BSM particles should only have a small coupling to SM particles, which would explain their small contribution to physical observables. The most prominent examples are dark photons, axions, or axion-like particles (ALPs).  The axion has been proposed as a dynamical solution to the strong CP problem \cite{Peccei:1977hh,Peccei:1977ur,Wilczek:1977md,Weinberg:1977ma}, i.e., the ``naturalness'' problem of the small QCD $\theta$ parameter. It is introduced as the Nambu-Goldstone boson associated with a spontaneously broken additional global $U(1)_\text{PQ}$ symmetry of the SM Lagrangian. The modified SM Lagrangian reads
\begin{align}
 \cL_\text{SM}^\text{eff.} &= \cL_\text{SM}+ \cL_\text{int}[\partial^\mu a_\text{phys.}/f_a; \psi]\\
 &-\frac{1}{2} \partial^\mu a_\text{phys.}\partial_\mu a_\text{phys.}-\frac{m_a^2}{2}\,a_\text{phys.}^2+\frac{a_\text{phys.}}{f_a}\zeta \frac{g_s^2}{32\pi^2} \tilde{G}_{\alpha\beta} G^{\alpha\beta}\nonumber,
\end{align}
where $a_\text{phys.}=a-\langle a \rangle$ is the physical axion field with mass $m_a$,
and $f_a$ is the $U(1)_\text{PQ}$ symmetry breaking scale. The axion is a pseudoscalar
that couples derivatively to any field $\psi$. 
In addition, because of the chiral anomaly of the $U(1)_\text{PQ}$ current, it directly
couples to the gluon density, where $\zeta$ is a model-dependent parameter. The minimum of the effective potential occurs at the axion vacuum expectation value $\langle a \rangle = - \theta f_a / \zeta$, which leads to a cancellation of the CP
violating QCD $\theta$ term and dynamically solves the strong CP problem. The defining characteristic of the axion,
distinguishing it from an ALP, is $m_a f_a \sim m_\pi f_\pi$. This follows from mixing of
the axion with the light $\pi$ and $\eta$ mesons.

In the following, we will be mainly concerned with heavy BSM particles. In
\secref{th:secoverviewSM}, we described how the  $W$ and $Z$ bosons can be integrated out
in an EFT approach. Similarly, whatever BSM physics there is, as long as it respects QED
and QCD gauge symmetry and involves degrees of freedom with a `large' mass
scale $\Lambda$, it can be integrated out and its effects will be
encoded in Wilson coefficients of gauge-invariant higher-dimensional
operators.  Operators that were absent in the SM case might now be
generated. Thus, we are led to write down the most general
relativistic Lagrangian that respects electromagnetic $U(1)_\text{em}$
and strong $SU(3)_c$ gauge invariance and obtain a general low-energy
effective field theory (LEFT)
\begin{align}
  \label{th:LEFT}
  \cL_\text{LEFT} =  \cL_\text{QED+QCD}
                 + \frac{1}{\Lambda} \sum_i C^{(5)}_i O^{(5)}_i
                 + \frac{1}{\Lambda^2} \sum_j C^{(6)}_j O^{(6)}_j
                 + \ldots 
\end{align}
Here $\Lambda$ is the scale of physics that is not dynamically
described by the degrees of freedom present in $\cL_\text{LEFT}$. If
we include all charged leptons and all quarks apart from the top in
$\cL_\text{LEFT}$, the scale $\Lambda$ is assumed to be larger than
the mass of the $b$ quark but not larger than the electroweak scale
$m_W$. The sums $i$ and $j$ run over all possible operators of
dimension 5 and 6, respectively. Typically, operators
  of dimension larger than 6 are neglected. $O^{(5)}$ and $O^{(6)}$
denote the operators, $C^{(5)}$ and $C^{(6)}$ are the corresponding
Wilson coefficients. Operators that are related through Fierz
identities or those that can be eliminated through equations of motion
are not included. Naturally, the choice of the operator basis is not
unique, but a complete basis up to dimension 6 can be found
in~\cite{Jenkins:2017jig}.

The Lagrangian \eqref{th:LEFT} provides a consistent quantum-field
theoretical framework to relate low-energy measurements to the
determination of parameters of the SM and constraints on BSM
physics. 
%\textcolor{blue}{
Many different routes have been taken to generically parametrize
low-energy observables and measuring or constraining the associated
parameters. The prime example is the Michel decay, where an analysis
with initially a single parameter~\cite{Michel:1949qe} was generalized
and written in terms of parameters related to scalar, vector and
tensor contact
interactions\hyperlink{ht:section06}{\footnote{\label{section06}
    Section 6: Muon decay \cite{section06}.}}~\cite{Fetscher:1986uj}. A
similar effort has been made for cLFV decays $\mu\to e\gamma$ and
$\mu\to eee$ considering lepton-flavor-violating contact
interactions~\cite{Kuno:1999jp}.

At first sight this is very similar to constraining the Wilson
coefficients of \eqref{th:LEFT}. Indeed, the bulk of the operators of
\eqref{th:LEFT} are also scalar, vector and tensor interactions.
However, the Wilson coefficients are well-defined couplings of a
quantum field theory. In particular, typically they run and mix under
renormalization-group evolution (RGE). If a low-energy observable is
expressed in terms of Wilson coefficients, they are understood to be
evaluated at the low scale, $C_i^{(n)}(m_\mu)$. On the other hand, to
relate the Wilson coefficients of the EFT to a BSM model, the heavy
degrees of freedom of the latter have to be integrated out. This
yields the Wilson coefficients at the high scale,
$C_i^{(n)}(\Lambda)$. Including RGE of
$C_i^{(n)}(\Lambda)$ to $C_i^{(n)}(m_\mu)$ is not in the first
instance about increasing precision, but to include qualitatively new
effects through mixing. This has a profound impact on using low-energy
measurements to constrain BSM models.
%}

Of course, it is also possible that BSM physics appears only at a scale much
larger than $m_W$. If this is the case, in a first step another
effective theory has to be used, the SM effective field theory
(SMEFT). This is a theory similar to \eqref{th:LEFT}, but with all
fields and symmetries of the SM. It contains all operators
$\mathcal{O}_i^{(n)}$ expressed in terms of the SM gauge fields, the
Higgs doublet, as well as left-handed doublet and right-handed singlet
fermion fields that respect the SM gauge symmetry $SU(3)_c\times SU(2)_L
\times U(1)_Y$,
\begin{align}
  \label{th:SMEFT}
  \cL_\text{SMEFT} =  \cL_\text{SM} 
     + \frac{1}{\Lambda} \big( \mathcal{C}^{(5)} \mathcal{O}^{(5)} + \mbox{h.c.} \big)
     + \frac{1}{\Lambda^2} \sum_j \mathcal{C}^{(6)}_j \mathcal{O}^{(6)}_j
     + \ldots
\end{align}
SMEFT has only one dimension-5 operator $\mathcal{O}^{(5)}$ (and its
Hermitian conjugate). This is the Weinberg
operator~\cite{Weinberg:1979sa} that is associated with neutrino
masses. At dimension 6 there are numerous operators, some of which
violate baryon number. As for $\cL_\text{LEFT}$ different bases are
possible, but the so-called Warsaw basis~\cite{Grzadkowski:2010es} is
used frequently.

%\textcolor{gray}{
In the case $\Lambda\gg m_W$ the input of the BSM model is given
through Wilson coefficients $\mathcal{C}_i^{(n)}(\Lambda)$. Then, the
RGE is used to obtain $\mathcal{C}_i^{(n)}(m_W)$. In a next step,
SMEFT is matched to LEFT at the electroweak scale. This means that
$C_i^{(n)}(m_W)$ are expressed in terms of $\mathcal{C}_i^{(n)}(m_W)$.
Finally, the Wilson coefficients of LEFT, $C_i^{(n)}(m_W)$, are run
with the RGE of LEFT from the scale $m_W$ to the low scale $m_\mu$,
and we are ready to express physical low-energy observables.
%}  
The complete dimension-6 RGEs of SMEFT and LEFT, and the matching
equations between the two EFTs are known at one
loop~\cite{Jenkins:2013zja,Jenkins:2013wua,
  Alonso:2013hga,Jenkins:2017dyc,Dekens:2019ept}, 
whereas beyond only partial results are known. 

Now that we have a framework that incorporates the effects of the full
SM and potential BSM physics on low-energy observables, we can return
to our starting point, the matrix elements of the electromagnetic
currents. Moving from \eqref{th:LagQEDQCD} to \eqref{th:LEFT} leads to
a generalization of \eqref{th:JQED}, \eqref{th:leptonJem}, and
\eqref{th:nucleonJem}. In particular, the current itself is modified
and includes additional terms from the dimension-5 dipole
operators. The most general expression for a vector current depending
on $p_1$ and $p_2$ can be written as combination of six possible
structures: $\gamma^\alpha$, $\gamma^\alpha \gamma_5$, $q^\alpha$,
$q^\alpha\gamma_5$, $q_\beta\sigma^{\alpha\beta}$ and
$q_\beta\sigma^{\alpha\beta}\gamma_5 $. Replacing $q=p_2-p_1$ by
$p_2+p_1$ does not lead to new independent structures, as can be shown
by using the Dirac equation. Since the electromagnetic current is
conserved $\partial_\alpha J_\text{em}^\alpha = 0$ only four terms
remain and we get
\begin{align}
  \label{th:genJem}
  \langle f(p_2)| J_\text{em}^\alpha |f(p_1)\rangle
   =  \bar{u}(p_2,m_f)\Big(& F_1^{(f)}(q^2)\, \gamma^\alpha
  + \big(F_2^{(f)}(q^2) -i\,\gamma_5  F_3^{(f)}(q^2) \big)\,
       \frac{i\, \sigma^{\alpha\beta}  q_\beta}{2\,m_f} \\
 \nonumber
       +\ & F_4^{(f)}(q^2)\, \frac{1}{m_f^2}
         \big(q^2 \gamma^\alpha - 2 m_f q^\alpha\big)\,\gamma_5 \Big)\,
  u(p_1,m_f) \, .
\end{align}
The
CP-violating form factor $F_3$ is associated with the EDM of the lepton $d_f$ through
\begin{align}
\label{th:F3df}
  d_f  = \frac{e F_3^{(f)}(0)}{2 m_f} \, .
\end{align} 
In the SM, $d_f$ starts to receive contributions at three loops for
quarks~\cite{Czarnecki:1997bu} and at four loops for
leptons~\cite{Pospelov:1991zt}, induced by the CP violation in the CKM
matrix. For protons and neutrons there is an additional source for an
EDM~\cite{Pospelov:1999ha} through the CP-violating $\theta$ term in
QCD
\begin{align}
\label{thetaterm}
	\cL_\text{QCD} \supset
\frac{g_s^2 \theta}{32\pi^2} \tilde{G}_{\alpha\beta} G^{\alpha\beta} \, ,
\end{align}
which we have neglected in \eqref{th:LagQEDQCD}.  This term has to be included as it respects
$SU(3)_c$ gauge invariance. Even though it can
be written as a total derivative and, so does not affect the
classical equations of motion, the $\theta$ term does have effects at
the quantum level. Thus strong interactions seem to violate
CP. However, due to experimental constraints on the neutron EDM, we
know that the $\theta$ parameter is extremely small, see
\secref{th:secn}. The lack of an explanation for this smallness is
referred to as the strong CP problem.  In generic BSM models, one
usually expects much larger CP-violating
effects~\cite{Pospelov:2005pr, Engel:2013lsa}. The parity-violating
anapole form factor $F_4$ is also induced due to weak interactions of
the SM, or potentially through BSM effects. However, it is not an
observable by itself~\cite{Musolf:1990sa}.

As mentioned above, matrix elements of the weak charged current
$J_\text{cc}^\alpha$ also play an important role. It gives rise to
non-vanishing matrix elements between different particles of
left-handed $SU(2)$ doublets, such as $(\nu_\ell,\ell)$ or $(p,n)$.
The former leads to muon decay, whereas the latter for example to beta
decay, or quasi-elastic scattering $\ell\,p\to \nu_\ell\,n$. In this
case, all six structures appear and setting $m_p=m_n\equiv m_N$ we
have
\begin{align}
  \label{th:matelJW}
   \langle p(p_2)| J_\text{cc}^\alpha |n(p_1)\rangle
   &= \bar{u}(p_2,m_N)\left( F_1^{(pn)}(q^2)\, \gamma^\alpha
   + F_2^{(pn)}(q^2) \frac{i\, \sigma^{\alpha\beta}  q_\beta}{2\,m_N}
   + F_A^{(pn)}(q^2)\, \gamma^\alpha \gamma_5\right.\\
   \nonumber
   &
 \left.  + F_P^{(pn)}(q^2) \frac{q^\alpha \gamma_5}{2\,m_N}
   + F_S^{(pn)}(q^2) \frac{q^\alpha}{m_N}
   + F_T^{(pn)}(q^2) \frac{i\, \sigma^{\alpha\beta}  q_\beta \gamma_5}{2\,m_N}
   \right) u(p_1,m_N) \, .
\end{align}
The scalar and tensor form factors $F_S$ and $F_T$ are referred to as
second-class currents and often are omitted. However, we will return
to them in \secref{th:secn} in connection with the nucleon $\beta^-$
decay, see \eqref{LeeYangbetadecay}, which can be related to
$F_{S,T}^{(pn)}$ and $F_{S,T}^{(\nu_ee^-)}$.  The axial-vector and
  the pseudoscalar form factors, $F_A^{(pn)}$, and $F_P^{(pn)}$ are
  related to often used couplings as
\begin{align}
\label{th:defgap}
  g_A&\equiv F_A^{(pn)}(0), & 
  \bar{g}_A&\equiv F_A^{(pn)}(q_0^2), &
  \bar{g}_P&\equiv \frac{m_\mu}{m_N} F_P^{(pn)}(q_0^2),
\end{align}
where $q_0^2 = -0.88\,m_\mu^2$ is the momentum transfer of $\mu^-$
capture on the proton, neglecting binding energies.

\subsubsection{Hadronic effects}\label{th:secoverviewHAD}

Not only the Wilson coefficients of the EFTs are subject to
RGEs and thus scale dependent, but also the gauge couplings $\alpha =
e^2/(4\pi)$ and $\alpha_s=g_s^2/(4\pi)$ in \eqref{th:LagQEDQCD}. Both
depend on the energy of the phenomenon they are used to describe, but
while $\alpha(Q^2)$ decreases towards $\alpha(0)\sim 1/137$, the
strong coupling $\alpha_s(Q^2)$
%, which describes the interaction between quarks and gluons, 
increases as we go to lower energies.
For energy scales below a couple
of GeV, a perturbative expansion in $\alpha_s$ no longer works --- the
relevant degrees of freedom related to the strong interactions at low
energies are not quarks and gluons, but light hadrons.  Once more, EFTs
come to the rescue, in this case chiral perturbation theory
(\chpt{})~\cite{Weinberg:1978kz,Gasser:1983yg,Leutwyler:1993iq}. As
for all EFTs, the first step is to identify the relevant degrees of
freedom in the energy range of interest. The second is to write down
the most general Lagrangian for these degrees of freedom that is
compatible with the symmetries of the underlying theory. For the
strong interactions the answer to the first question is related to the
phenomenon of spontaneous chiral symmetry breaking, which generates
Goldstone bosons, the only massless particles of strong
interactions. Actually in the spectrum of QCD there are no massless
particles, but a triplet of very light pseudoscalars, the pions
$\vec{\pi}=(\pi^+, \pi^0, \pi^-)$. The fact that they are not exactly
massless is well understood and due to the presence of an explicit,
but small, chiral symmetry breaking term in the QCD Lagrangian: the
quark mass term. In the limit of zero up and down quark masses, i.e.,
$m_d = m_u = 0$, the three pions become massless, and since there are
no other mechanisms to generate massless particles in QCD in the
chiral limit, these are the only relevant degrees of freedom at low
energy.

The rules to write down an effective Lagrangian for Goldstone bosons
are well known. Goldstone bosons transform nonlinearly under the
symmetry of the underlying theory, which leads to a non-renormalizable
Lagrangian containing only derivative couplings. Symmetry constrains
their interaction to become weaker as one lowers the energy. How to
include an explicit symmetry breaking is also well known. The symmetry
breaking parameters are promoted to spurions, fields with given
transformation laws, and the effective Lagrangian must include these
fields too and still satisfy the requirement of being invariant under
symmetry transformations. In the case of QCD, in addition to
derivative couplings, it is also possible to have couplings
proportional to the quark masses $m_{u,d}$. Clearly, there are
infinitely many such terms and the Lagrangian only becomes useful with
an organizing principle. Since this is a low-energy EFT, we count
powers of energy or momenta as small, and since it is relativistic,
they come in even powers. The smallest possible number is two, then
four, six and so on. Quark masses (or explicit symmetry breaking in
general) also count as small, but there is no unique choice concerning
the relative importance of powers of quark masses and derivatives. The
standard one is $m \sim p^2$. According to this choice the
lowest-order Lagrangian contains all possible terms with two powers of
derivatives or one power of quark masses and it turns out that there
are only two:
\begin{align}
   \mathcal{L}_\mathrm{\chi PT}&=\mathcal{L}_2+\mathcal{L}_4+\mathcal{L}_6+\ldots \,, \quad \mathcal{L}_2=\frac{F^2}{4}\langle u_\mu u^\mu + \chi_+ \rangle \, ,\label{LChPT}
\end{align}
where $u_\mu=iu^\dagger \partial_\mu U u^\dagger$, $\chi_+=u^\dagger \chi u^\dagger+u \chi^\dagger u$, and 
\begin{align}
U= u u &=\exp\left(i \phi/F\right) \, , \quad \phi= \pi^a \tau_a \, , \quad \chi = 2 B \, \mbox{diag}(m_u,m_d)   \, ,
\end{align}
with $\pi^a$ the triplet of pion fields and $\tau_a$ the Pauli matrices. The low-energy constant (LEC) $F$ is related to  the pion decay constant
\begin{align}
\label{th:pideccst}
\langle 0| (J_A^a)_\mu(0) | \pi^b(p) \rangle&= i \delta^{ab} F_\pi p_\mu \, , \quad F_\pi=F\left(1+\mathcal{O}(m_q)\right)\, , 
\end{align}
with $(J_A^a)_\mu$ the isospin-triplet axial current. The second LEC $B$ is defined through the quark condensate in the chiral limit,
\begin{align}
B=-\frac{\langle0\vert \bar{u}u\vert0\rangle}{F^2}
=-\frac{\langle0\vert \bar{d}d\vert0\rangle}{F^2} \, ,
\end{align}
and also relates the pion mass to the quark mass according to the Gell-Mann--Oakes--Renner relation \cite{Gell-Mann:1968hlm}
\begin{align}
   m_\pi^2=2 B \hat{m}\left(1+\mathcal{O}(m_q)\right) \, , 
\end{align}
with
$\hat{m}=(m_u+m_d)/2$.
 Calculating tree-level diagrams with
$\mathcal{L}_2$ gives a leading-order (LO) result. Going to
next-to-leading order (NLO) requires calculating one-loop diagrams
with vertices only from $\mathcal{L}_2$ and tree-level diagrams with
one vertex from
$\mathcal{L}_4$~\cite{Weinberg:1979sa,Gasser:1983yg}. At
next-to-next-to leading order (NNLO) two-loop diagrams with vertices
only from $\mathcal{L}_2$, one-loop diagrams with one vertex from
$\mathcal{L}_4$ and tree-level diagrams with two vertices from
$\mathcal{L}_4$ or one from $\mathcal{L}_6$
contribute~\cite{Fearing:1994ga,Bijnens:1999hw,Bijnens:1999sh}, and so
on.

The limit of validity of this EFT is given by the scale of chiral
symmetry breaking. In the expansion in powers of momenta and quark
masses that is generated by the effective Lagrangian above, the
relevant scale is represented by $\Lambda_\chi =4 \pi F_\pi \sim 1.2$
GeV. Physically it represents the scale at which degrees of freedom
other than Goldstone bosons get excited, such as the $\rho$, whose
mass $m_\rho\sim 0.77$ GeV is indeed close to $\Lambda_\chi$.

The same approach also works for other particles beyond the pions. In
the limit $m_s \to 0$ also the kaons and the eta become Goldstone
bosons and can be included in the formalism
above~\cite{Gasser:1984gg}. The field $\phi$ becomes a $3\times 3$
matrix containing the octet of Goldstone bosons $\phi=\phi^a
\lambda_a$, and $\chi$ has to be trivially extended to a diagonal
$3\times 3$ quark-mass matrix.

A less trivial extension concerns the baryon
sector~\cite{Gasser:1987rb,Jenkins:1990jv,Bernard:1992qa,Becher:1999he}. At
first sight this would seem impossible, since the mass of the nucleons
is close to $\Lambda_\chi$. But the baryon number $n_B$ is conserved
in strong interactions and one can split the spectrum in separated
sectors, labeled by $n_B$. Quantities like the nucleon masses, their
form factors, or their scattering amplitude with a pion (or any other
Goldstone boson(s)) all belong to the sector $n_B=1$ and can also be
studied with the help of the chiral expansion. In this case this
represents an expansion in powers of momenta and quark masses around
the ground-state energy, which in this sector is equal to the mass of
the nucleon $m_N$, rather than zero.

From the point of view of their transformation properties, nucleons
are spin-1/2 as well as isospin-1/2 particles, and transform linearly
under chiral transformations. In particular the fact that they are
spin-1/2 particles has an important consequence as the expansion of
the Lagrangian in powers of momenta (derivatives) contains both even
and odd powers
\begin{align}
    \mathcal{L}_N&=\mathcal{L}_1+\mathcal{L}_2+\mathcal{L}_3+\ldots
\end{align}
The leading-order Lagrangian looks as follows
\begin{align}
    \mathcal{L}_1&=\bar{N}(i / \!\!\!\!D-m)N+ \frac{1}{2}g_A \bar{N} / \!\!\!\!u \gamma_5 N ,
\end{align}
with the covariant derivative defined as
\begin{align}
    D_\mu&=\partial_\mu+\Gamma_\mu\, , \quad \Gamma_\mu=\frac{1}{2} [u^\dagger, \partial_\mu u] \; ,
\end{align}
and $\bar{N}=(\bar p,\bar n)$ the isospin doublet containing the Dirac
spinors of the proton and neutron. The parameters $m$ and $g_A$
represent the mass and the axial coupling of the nucleon in the chiral
limit, respectively. Note that the chiral symmetry imposes the
presence of the pion field both in the covariant derivative as well as
in the coupling to the nucleon axial current. From this follows the
famous Golberger-Treiman relation~\cite{Goldberger:1958zz}
\begin{align}
    g_{\pi N}&= \frac{g_A m_N}{F_\pi} ,
\end{align}
between the pion-nucleon coupling constant $g_{\pi N}$ (whose square
is the residue of the nucleon pole in the $\pi N$ scattering
amplitude), the physical nucleon mass, and the axial coupling.

The low-energy description of the strong-interaction effects in terms
of \chpt{} cannot only be formulated for pure QCD as the underlying
theory. While QED effects can be included in terms of explicit
low-energy degrees of freedom, the chiral realization of
higher-dimensional operators again is based on the external-field and
spurion technique. Traditionally, this has been done to include
weak-interaction effects and it can be generalized to include BSM
effects encoded in the LEFT Lagrangian~\eqref{th:LEFT}.

\subsection{The muon} \label{th:secmu}

The muon is a fundamental lepton similar to the electron, however with
a much larger mass, $m_\mu\simeq 105.66 \MeV$.  It is unstable and
predominantly decays through the Michel process
\begin{align}
  \label{th:michel}
  \mu\to e \nu \bar\nu \,,
\end{align}
which leads\hyperlink{ht:section16}{\footnote{\label{section16}
    Section 16: MuLan \cite{section16}.}} to a lifetime of about
$\tau_\mu \simeq 2.2\,\mu\text{s}$.  As discussed in the context of
\eqref{th:matelJW} the decay is mediated by the charged current
$J^\alpha_\text{cc}$, leading to a non-vanishing current-current
interaction $\langle \nu_\mu|J^\alpha_\text{cc}|\mu\rangle \, \langle
e|(J_\text{cc})_\alpha^\dagger|\nu_e\rangle$. From an EFT point of
view this corresponds to a four-fermion operator $(\bar\nu_\mu
\gamma^\alpha P_L \mu) (\bar{e} \gamma_\alpha P_L {\nu_e})$ and its
Hermitian conjugate. For computational reasons it is more convenient
to work with the Fierz transform of this operator. This results in the
Fermi theory, an EFT defined through the Lagrangian
\begin{align}
  \label{th:LagMu}
  \cL_\text{Fermi} = -\frac{4\, G_F}{\sqrt2} 
\left( \bar\nu_\mu \gamma_\alpha P_L {\nu_e} \right)
\left( \bar{e} \gamma^\alpha P_L \mu \right)  
+ \text{h.c.} 
+ \cL_{\text{QED}+\text{QCD}} \, ,
\end{align}
where it is implicitly assumed that only light quarks are included in $\cL_\text{QCD}$.
The first term on the r.h.s. of \eqref{th:LagMu} corresponds to the
operator $[O^{V,LL}_{\nu\ell}]_{2112}$ as introduced in
\eqref{th:opV4f}.  Its Wilson coefficient, $4\, G_F/\sqrt2$, has the
special property that it does not get
renormalized~\cite{Berman:196220}. Thus, the Lagrangian
\eqref{th:LagMu} can be used to consistently compute at leading order
in $G_F$ but to all orders in the electromagnetic coupling
$\alpha$. Only the usual QED renormalization procedure has to be
applied. As an example, the lifetime of the muon can be expressed as
\begin{align}
  \label{th:Mulife}
  \frac{1}{\tau_\mu} \equiv \Gamma_\mu = \Gamma_0 \big(1 + \Delta q \big)
    = \frac{G_F^2 m_\mu^5}{192\, \pi^3} \big(1 + \Delta q \big) ,
\end{align}
where $\Delta q$ contains all corrections to $\Gamma_0$ (the tree-level
result for massless electrons) that are induced by \eqref{th:LagMu}. This
includes electron-mass effects, higher-order QED corrections, as well as
hadronic corrections. While the former two can be computed in perturbation
theory, the latter are more delicate. As mentioned above, QCD is
non-perturbative at scales typical for
muonic processes, $q^2\sim m_\mu^2$. Thus, the hadronic contributions
have to be determined by other means.  This is often the leading
theoretical uncertainty. The fact that such corrections for muonic
processes enter only at NNLO makes the muon a rather clean laboratory
for precision physics. Typically, $\cL_\text{QED}$ contains muon
and electron fields, but the inclusion of $\tau$ leptons is
straightforward, as is the inclusion of heavy quarks in $\cL_\text{QCD}$.

The corrections $\Delta q$ are known at NNLO with full electron mass
dependence~\cite{vanRitbergen:1998yd,  Anastasiou:2005pn, Pak:2008qt,
Engel:2019nfw}. Thus, with a precision measurement of the muon lifetime,
the Wilson coefficient in \eqref{th:LagMu}, or equivalently $G_F$, can be
determined extremely precisely. This, in turn, is an important input for
electroweak precision tests. In fact, $G_F$ can be related to $m_W$ and
$m_Z$ through
\begin{align}
  \label{th:GFMW}
  \frac{4\, G_F}{\sqrt2} &=
  \frac{g^2}{2 m_W^2} \big(1 + \Delta r \big) =
  \frac{2\pi\, \alpha}{\sin^2\theta_W\, m_W^2} \big(1 + \Delta r \big) ,
\end{align}
where (in the on-shell scheme) $\sin^2\theta_W=1-m_W^2/m_Z^2$.  
The SM corrections $\Delta r$ contain (partially hadronic) fermion loop contributions to
the charge renormalization. Additional contributions depend also on the top and Higgs
mass. This makes $G_F$ a decisive input for SM consistency checks. As mentioned
in~\cite{section16} only the availability of the NNLO result~\cite{vanRitbergen:1998yd}
allowed for a full exploitation of the experimental results. 

While SM corrections are crucial for the electroweak
precision tests the tree-level matching of the SM to the Fermi theory
yields the matching condition \eqref{th:GFMW} with $\Delta r\to 0$ that is
used in \eqref{th:LagMu}. Furthermore, terms of order $q^2/m_W^2$ relative
to the four-fermion interaction are also neglected in \eqref{th:LagMu} and
typically in \eqref{th:LEFT}. In the literature \eqref{th:Mulife} is
often written with an additional factor $(1 + 3/5\, (m_\mu/m_W)^2)$
which results in a $10^{-6}$ correction. Within the EFT, such
corrections are reproduced by dimension-8 operators, which are missing
in \eqref{th:LagMu}. There are also numerous dimension-6 operators
generated by the SM that are not included in \eqref{th:LagMu}.
The corresponding Wilson coefficients are related to the general
parametrization of muon decay parameters.\footref{section06}

Apart from the Michel decay, two further SM decay processes are of
interest; the radiative and rare decays
\begin{align}
  \label{th:radrare}
  \mu&\to e \nu \bar\nu \gamma\, ,&
  \mu&\to e \nu \bar\nu e^+ e^-\, .
\end{align}
In order to be well defined and to avoid infrared singularities, the
branching ratio for the radiative decay must be defined requiring a
minimal energy of the photon. For $E_\gamma > 10\,\MeV$ we have
$B(\mu\to e \nu \bar\nu \gamma )\sim 1.3\times 10^{-2}$. For the rare
decay the branching ratio is \begingroup
	\fontsize{10.5pt}{12}\selectfont
	$B(\mu\to e \nu \bar\nu ee)\sim 3.6\times
10^{-5}$.
	\endgroup A fully differential NLO description of these processes in
the Fermi theory \eqref{th:LagMu} is available~\cite{Fael:2015gua,
  Pruna:2017upz, Pruna:2016spf, Fael:2016yle}. Depending on the cuts
that are applied, the NLO QED corrections can be
sizeable. Experimental information on the branching ratio of the
radiative decay has been obtained by MEG~\cite{Adam:2013gfn} and
PiBeta~\cite{Pocanic:2014mya}.

A particularly attractive feature of particle physics with muons is
the study of cLFV decays. There are three "golden" channels
\begin{align}
  \label{th:cLFVgold}
   \mu&\to e \gamma\, , &
   \mu&\to eee , &
   \mu^-\, \ce{_Z^A}N &\to e^-\, \ce{_Z^A}N \, .
\end{align}
PSI has a long tradition in corresponding experimental
searches.\hyperlink{ht:section07}{\footnote{\label{section07} Section
    7: SINDRUM \cite{section07}.}\footnote{\label{section08} Section 8:
    SINDRUM II \cite{section08}.}\footnote{\label{section19} Section
    19: MEG \cite{section19}.}\footnote{\label{section20} Section 20:
    Mu3e \cite{section20}.}} For the first two processes typically
$\mu^+$ are used, whereas $\mu^-$ must be used for muon conversion
in the field of a nucleus $\ce{_Z^A}N$ with atomic number Z and mass
number A. In the SM (with non-vanishing neutrino masses) the branching
ratios for these processes are smaller than $10^{-50}$, but not
zero~\cite{Petcov:1976ff}. Hence, from a theory point of view there is
nothing sacred about lepton flavor. As we know that it is not
conserved, it is very natural to expect much larger cLFV branching
ratios in BSM than in the SM.  In fact, generic extensions of the SM
do typically lead to large cLFV rates and suppressing them requires
additional tuning or model-building efforts.

%\textcolor{gray}{
To extract constraints on BSM physics from limits on the
branching ratios of the processes \eqref{th:cLFVgold}, they are
computed in $\cL_\text{LEFT}$, typically at tree level. For $\mu\to e
\gamma$ the dipole operator $[O^D_{\ell\gamma}]_{21}$~\eqref{th:opD}
enters. Thus we get a limit on the corresponding Wilson coefficient
at the low scale $[C^D_{\ell\gamma}]_{21}(m_\mu)$. In a next step, the
RGE is used to convert this to limits for the Wilson coefficients at
the high scale, $C_i(\Lambda)$.  Some scalar four-fermion interactions
mix at NLO whereas vector four-fermion interactions enter at
NNLO. Nevertheless, this results in very stringent limits on contact
interactions induced by BSM physics. They have to be combined with
limits from $\mu\to eee$ and muon conversion, where contact
interactions already appear at leading order. Using as many operators
as possible in connection with RGE maximizes the information that can
be obtained from low-energy observables.
%}

These computations can be made ~\cite{Crivellin:2017rmk} for $\mu\to e \gamma$ and $\mu\to eee$ using standard perturbative methods with
the Lagrangian \eqref{th:LEFT}, although for some contributions,
non-perturbative effects play a role~\cite{Dekens:2018pbu}.  However,
additional input is required for muon conversion. First, the nuclear
matrix elements $\langle \ce{_Z^A}N|J|\ce{_Z^A}N\rangle$ for vector and
scalar currents/operators are required. The former can be obtained
trivially through current conversion, but the latter need input from
lattice QCD or $\chi$PT. Second, the overlap integrals of the lepton
wave function with the nucleus are required~\cite{Kitano:2002mt}. In
principle different target nuclei provide different limits on the
various coefficients, but in practice the model discriminating power
is limited~\cite{Cirigliano:2009bz}. A further complication is due to
background from the decay in orbit (DIO). This is the Michel decay of
the $\mu^-$ bound in the nucleus
\begin{align}
  \label{th:DIO}
  \mu^-\, \ce{_Z^A}N \to e^-\, \nu_\mu \bar\nu_e\, \ce{_Z^A}N\, .
\end{align}
Due to nuclear recoil effects the energy spectrum of the
electron has a tail up to $m_\mu$, the energy of the signal for the electron
from muon conversion. Thus DIO has to be studied as a background
process~\cite{Czarnecki:2014cxa}. 

So far the nucleus has acted only as a spectator. The only nuclear physics
that was required is the nuclear matrix element. For completeness we
mention here two processes relevant to muon conversion, where the
nuclear physics is much more involved. When the $\mu^-$ is bound to
the nucleus, it quickly cascades to the $1S$ ground state. Then it
might undergo muon capture
\begin{align}
  \label{th:mucapN}
  \mu^-\, \ce{_Z^A}N \to \nu_\mu\, \ce{_{Z-1}^A}N \, 
\end{align}
before it decays. The corresponding nuclear matrix element $\langle
\ce{_{Z-1}^A}N|(J_\text{cc}^\alpha)^\dagger|\ce{_Z^A}N\rangle$ is an extended
version of \eqref{th:matelJW}. It depends on the details of $\ce{_Z^A}N$
and is not easily accessible with theoretical methods. We will return
to muon capture in Section~\ref{th:secp}.

%\textcolor{green}{Another cLFV process that can in principle be considered is the
%`wrong' muon conversion~\cite{Geib:2016atx}
%\begin{align}
%  \label{th:muconvX}
%  \mu^-\, \ce{_Z^A}N &\to e^+\, \ce{_{Z-2}^A}N \, .
%\end{align}
%This process actually even violates lepton number conservation. From
%a theoretical point of view, these processes are much harder to
%describe, since the matrix element $\langle \ce{_{Z-2}^A}N|J|\ce{_Z^A}N\rangle$
%is highly non-trivial. This is a problem akin to neutrinoless double
%beta decay. }

The muon can not only form bound states with a nucleus, but also with
an electron. Muonium, $M=(\mu^+ e^-)$, is a bound state like hydrogen,
but with the proton replaced by a positive muon. As the latter
is a pointlike fermion, muonium is an excellent laboratory for QED
tests, and for a precise determination of the muon
mass.\hyperlink{ht:section29}{\footnote{\label{section29} Section 29: MSpec, Mu-Mass
  \cite{section29}.}} As the muonium mass is dominated by antimatter, $M$ is
also an interesting option to study experimentally gravity of
antimatter~\cite{Antognini:2018nhb}. In addition, muonium-antimuonium
oscillations
\begin{align}
  \label{th:muonium}
  M=(\mu^+ e^-) & \leftrightarrow  \bar{M}=(\mu^- e^+) \, ,
\end{align}
which are forbidden in the SM, are another channel to scrutinize BSM
physics.\hyperlink{ht:section09}{\footnote{\label{section09} Section 9: MACS \cite{section09}.}}
A bound state of two muons, true muonium $(\mu^+ \mu^-)$, is
unfortunately, not experimentally accessible in the foreseeable future.

Two further properties of the muon that are of utmost importance are
the AMM~\eqref{th:F2al} and EDM~\eqref{th:F3df}. The motivation to
study them in detail is again driven by the desire to test the SM. For
the AMM very precise measurements are confronted with similarly
precise theoretical predictions~\cite{Aoyama:2020ynm}. At the time of
writing, there is an intriguing tension between SM theory and
experiment. For the EDM, the situation is similar to cLFV searches in
that the SM value is zero for practical experimental purposes. Hence,
experimental verification of a non-vanishing muon EDM is a clear
indication of BSM.  So far, these quantities have not been measured by
PSI experiments. However, future involvement, in particular for the
EDM, is being considered~\cite{Kirch:2020lbo}.

\subsection{The proton}\label{th:secp}

Like the electron and muon, the proton is a charged spin 1/2
fermion. However, because the proton is a bound state, the form factors
\eqref{th:nucleonJem} cannot be computed perturbatively simply using
$\cL_\text{QED+QCD}$.  Most information is obtained from experiment,
with additional input from lattice QCD and
$\chi$PT~\cite{Perdrisat:2006hj}.  From the charge and measurements of
the AMM we know $F_1^{(p)}(0)=1$ and $F_2^{(p)}(0)=\kappa_p\simeq
1.79$.

A quantity that has received a lot of attention in the past years is
the proton charge radius $r_E^{(p)}$. As discussed in the context of
\eqref{th:radius}, the radius can be extracted as the slope of
$G_E^{(p)}(q^2)$ at $q^2\to 0$. This can be determined by low-$q^2$
lepton-proton scattering with a careful $q^2\to 0$ extrapolation. An
alternative approach is to use spectroscopy of normal hydrogen or
better muonic hydrogen. The overlap of the lepton wave function with
the proton charge distribution impacts on the energy levels. Thus, a
precise measurement of different transition energies allows the extraction of
information on the proton radius. As the Bohr radius is proportional to
$1/m_\ell$, the effect in muonic atoms is considerably larger. This has
resulted in a very precise new determination of the proton
radius\hyperlink{ht:section21}{\footnote{\label{section21} Section 21: CREMA \cite{section21}.}}
and a new world average of $r_E^{(p)}\simeq 0.84$~fm. The disagreement
with earlier determinations of $r_E^{(p)}$ was referred to as proton
radius puzzle~\mbox{\cite{Pohl:2013yb, Carlson:2015jba}}, but the puzzle is
fading away~\cite{Hammer:2019uab}.

%\textcolor{gray}{
The CREMA collaboration\footref{section21} has
measured two transition frequencies for muonic hydrogen; the triplet
$E\left(2P_{3/2}^{F=2}\right)-E\left(2S_{1/2}^{F=1}\right)$ and singlet
$E\left(2P_{3/2}^{F=1}\right)-E\left(2S_{1/2}^{F=0}\right)$. From these two values and
theoretical input for the fine structure, it is possible to extract the
Lamb shift $E_L=E\left(2P_{1/2}\right)-E\left(2S_{1/2}\right)$ and the hyperfine splitting
$E_{HFS}=E\left(2S_{1/2}^{F=1}\right)-E\left(2S_{1/2}^{F=0}\right)$.
%} 
The discrepancy of the proton radius determination from muonic
hydrogen with earlier values initiated a flurry of activities to
revisit the theoretical calculations of the energy levels, as
summarized in~\cite{Antognini:2013jkc}. This involves radiative
corrections and recoil effects, which can in principle be computed in
perturbation theory.

In addition there are proton-structure effects, which are divided into
two categories: a) finite-size effects, which depend on the charge
$\rho_E$ and magnetic moment distribution $\rho_M$ of the proton,
i.e., the charges related to the form factors $G_E^{(p)}$ and
$G_M^{(p)}$, introduced in \eqref{th:FtoG}; b) polarizability effects.

The leading finite-size effect for $E_L$ is in fact proportional to
$\left(r_E^{(p)}\right)^2$ and it is precisely this effect that allows an accurate determination of $r_E^{(p)}$ from muonic hydrogen
spectroscopy to be made. There are also higher-order effects which have to be
included, most notably a contribution from the so-called third Zemach
moment
\begin{align}\label{th:friar}
\left(r_F^{(p)}\right)^3 &\equiv \frac{48}{\pi} \int_0^\infty \frac{dQ}{Q^4}
\bigg( \left[G_E^{(p)}(Q^2) \right]^2 -  1 +\frac{1}{3} \left[r_E^{(p)}\right]^2\, Q^2 \bigg)\, ,
\end{align}
where $r_F^{(p)}$ is referred to as Friar
radius.  This contribution is related to the elastic two-photon
exchange (TPE), where elastic refers to the fact that the intermediate
hadronic state is still a proton. The inelastic TPE, i.e., TPE where
the intermediate hadronic state is more complicated, is often referred
to as polarizability correction.

A similar distinction between perturbative and finite-size
contributions can be made for the hyperfine splitting $E_{HFS}$. In
this case, the leading finite-size effect is proportional to the
Zemach radius  $r_Z^{(p)}\simeq 1.0$~fm, a convolution of the charge
distribution with the magnetic moment distribution
\begin{align}
  \label{th:zemach}
  r_Z^{(p)}&\equiv \int d^3\vec{r}_1 \int d^3\vec{r}_2\
  \rho_E^{(p)}(\vec{r}_1) \rho_M^{(p)}(\vec{r}_2) |\vec{r}_1-\vec{r}_2|\, .
\end{align}
While the determination of the magnetic radius of the proton
$r_M^{(p)}\simeq 0.8$~fm was discussed less controversially, there is
also quite a spread in the values obtained from different
extractions~\cite{Alarcon:2020kcz}. This spread is typically
attributed to different treatment of TPE contributions.

%\textcolor{blue}{

The CREMA collaboration also investigated muonic deuterium
%\cite{Pohl1:2016xoo} 
and helium\footref{section21}
%\cite{crema:helium} 
and determined the corresponding charge radii. Measuring the charge radii
of higher $Z$ nuclei\hyperlink{ht:section22}{\footnote{\label{section22} Section 22: muX \cite{section22}.}}
provides crucial input for potential atomic parity violation
experiments.
%}

Returning to the proton, as mentioned above, studying lepton-proton
scattering at low $q^2$ is an important source to obtain information
on the proton form factors and, hence, the proton radius. At tree
level, which implies the one-photon approximation, this process is
described by the famous Rosenbluth formula
\begin{align}\label{th:rbluth}
  \frac{d\sigma}{d\Omega} &= \frac{\alpha^2}{4\,E^2_1\sin^4\theta_2}
   \frac{E_3}{E_1} \left(\frac{\left[G_E^{(p)}(q^2) \right]^2 + \tau\left[G_M^{(p)}(q^2) \right]^2}{1+\tau}
   \cos^2\theta_2  + 2 \tau \left[G_M^{(p)}(q^2) \right]^2 \sin^2\theta_2 \right) ,
\end{align}
in terms of $\tau=-q^2/(4 m^2_p)$, the scattering angle
$\theta=2\theta_2$, and the energies of the incoming and outgoing
leptons, $E_1$ and $E_3$, respectively. Using the standard dipole form $G_D(q^2)$
for the form factors gives a good fit to the experimental data:
\begin{align} \label{th:dipoleFF}
  G_E^{(p)}(q^2) &\simeq \frac{G_M^{(p)}(q^2)}{1+\kappa_p} \simeq G_D(q^2)=
  \frac{1}{(1-q^2/\Lambda^2)^2} \qquad
  \mbox{with}\quad \Lambda^2 = 0.71\,\mathrm{GeV}^2.
\end{align}
For very small $q^2$ the form factors deviate from \eqref{th:dipoleFF}
and --- coming back to the proton radius issue --- it is a delicate
problem to extract the slope of the form factors in the limit $q^2\to
0$ from scattering data.

Given the importance of lepton-proton scattering, there is a vast
literature on the computation of higher-order corrections to
\eqref{th:rbluth}. These corrections can be split into gauge
independent and finite subsets by separately considering radiative
corrections from the lepton line, radiation from the proton line, and
multi-photon exchange between the proton and electron.

A full NLO calculation, superseding earlier ones where various
approximations had been used, has been presented
in~\cite{Maximon_2000} and there are several Monte Carlo generators
with these corrections implemented~\cite{Gramolin:2014pva,
  Akushevich:2015toa}.  Corrections at NNLO due to radiation from the
electron line have also been computed~\cite{Bucoveanu:2018soy,
  Banerjee:2020rww}.  Due to the small mass of the lepton, these are
the dominant corrections, particularly for electron-proton
scattering. As for spectroscopy, from a theoretical point of view,
multi-photon exchange contributions between the lepton and proton are
the most difficult ones to handle. Accordingly, TPE contributions have
received a lot of attention, also including the inelastic parts, see
e.g.~\cite{Carlson:2007sp, Arrington:2011dn, Afanasev:2017gsk,
  Tomalak:2017shs}.

Traditionally, these experiments have been carried out with
electrons. The MUSE collaboration\hyperlink{ht:section23}{\footnote{\label{section23} Section
  23: MUSE \cite{section23}.}} proposes to measure $\ell\,p\to \ell\,p$
with $\ell\in\{e^\pm, \mu^\pm\}$. This offers the opportunity to
compare $e\,p$ and $\mu\,p$ scattering within the same experimental
setup. In addition, experimental information on TPE can be obtained by
measuring the difference between $\ell^+ p$ and $\ell^- p$ scattering.

To the best of our knowledge, the proton is a stable particle and in
all processes discussed so far, has been left intact. A low-energy
process that affects the proton much more dramatically is muon
capture, $\mu^-\,p\to n\nu_\mu$. This process can be described by the
transition matrix element \eqref{th:matelJW} as a current-current
interaction $\langle \nu_\mu| J_\text{cc}^\alpha |\mu\rangle\, \langle
n|(J_\text{cc})^\dagger_\alpha |p\rangle$. In fact, muon capture on
the proton as measured by MuCap\hyperlink{ht:section17}{\footnote{\label{section17} Section 17:
  MuCap \cite{section17}.}} gives valuable information on the
corresponding form factors, in particular $\bar{g}_P$
\eqref{th:defgap}~\cite{Hill:2017wgb}. The inverse process would be
related to neutrino-nucleon scattering. Muon capture
  on the deuterium has been investigated by
  MuSun.\hyperlink{ht:section18}{\footnote{\label{section18} Section 18: MuSun
    \cite{section18}.}}

\subsection{Nucleons and nuclei} \label{th:secnuc}

%\textcolor{gray}{
The proton and neutron together form an isospin doublet. They differ by
their isospin projection, $I_3=+\nicefrac{1}{2}$ and
$I_3=-\nicefrac{1}{2}$, and quark content, $uud$ and $udd$,
respectively.
%} 
The neutron's Dirac and Pauli form factors are normalized as
$F_1^{(n)}(0)=0$ and $F_2^{(n)}(0)=\kappa_n\simeq -1.91$. The former
differs from the proton form factor at zero momentum transfer,
$F_1^{(p)}(0)=1$, due to the vanishing charge of the
neutron. Therefore, the electric Sachs form factor of the neutron
cannot be approximated with a dipole form factor
\eqref{th:dipoleFF}. Instead, the Galster form factor could be used as
a simple parametrization \cite{Galster:1971kv}:
\begin{equation}
G_E^{(n)}(q^2)=\frac{q^2 \kappa_n}{4m_n^2-\eta q^2}G_D(q^2),
\end{equation}
with $\eta=5.6$. Since there are no free neutron targets, one has to
rely on scattering off light nuclei (e.g., $^2$H or $^3$He) to extract
the neutron form factors and polarizabilities. Thereby, few-nucleon
EFTs are needed to separate the neutron from proton and nuclear
effects.

As highlighted in the previous section, muonic atoms are sensitive to
the nuclear structure. The measurement of the muonic-hydrogen Lamb
shift by the CREMA collaboration\footref{section21} allowed the
extraction of the proton root-mean-square charge radius with
unprecedented precision. From the measured the Lamb shifts in $\mu$D,
$\mu^3$He$^+$ and $\mu^4$He$^+$
%\cite{Pohl1:2016xoo,Krauth2021} 
the deuteron, helion and $\alpha$-particle charge radii can be
extracted. In the future, the ground-state hyperfine splitting of
$\mu^3$He$^+$ shall be measured to extract the helion Zemach
radius. To extract the different nuclear radii, precise theory
predictions for the energy levels in muonic atoms are needed, see
theory summaries in
\cite{Krauth:2015nja,Franke:2017tpc,Diepold:2016cxv}. Among other
contributions, one needs the finite-size effects, through which the
different radii enter, and the polarizability effects. For the light
muonic atoms, not only the proton polarizability enters, but also the
polarizabilities of the neutron and the nucleus as a whole. Similar
complications arise when going from pionic hydrogen to pionic
deuterium\hyperlink{ht:section14}{\footnote{\label{section14} Section 14: Pionic hydrogen and
  deuterium \cite{section14}.}} or helium.\hyperlink{ht:section26}{\footnote{\label{section26}
  Section 26: Pionic helium \cite{section26}.}} The nuclear
polarizabilities are typically several orders of magnitude
larger than the nucleon polarizabilities, and thus, more
important. 
%\textcolor{blue}{
Take for instance the electric dipole polarizability,
$\alpha_{E1}^{(n)}=11.8(1.1)\times10^{-4}$ fm$^3$ \cite{Zyla:2020zbs}
and $\alpha_{E1}^{(d)}=0.6314(19)$ fm$^3$ \cite{Phillips:1999hh},
which describes the deformation of a composite particle in an external
electric field and gives a dominant contribution to the two-photon
exchange.
%} 
The nuclear polarizability effects can be calculated in a dispersion
relations framework \cite{Carlson:2013xea,Carlson:2016cii} or based on
nuclear potentials. For the latter, one distinguishes calculations
with phenomenological models \cite{Pachucki:2015uga} fit to
nucleon-nucleon scattering data, such as the AV18 potential
\cite{Wiringa:1994wb}, or with nucleon-nucleon interactions derived
from chiral EFT
\cite{Ji:2013oba,Hernandez:2014pwa,Dinur:2015vzv,Ji:2018ozm}. The
nucleon-structure contributions are often deduced by rescaling the
proton-structure contributions to $\mu$H. Take, for example, the
nucleon-polarizability contribution
  \begin{equation}
 \delta^\text{N}_\mathrm{pol}(\mu\text{A})=(\text{N}+\text{Z})\left[\text{Z}m_r(\mu\text{A})/m_r(\mu\text{H})\right]^3 \delta^\text{N}_\mathrm{pol}(\mu\text{H}),
 \end{equation}
where $m_r$ is the reduced mass of the muonic atom and Z, N, A are the
numbers of protons, neutrons and nucleons in the nucleus.
 %The elastic two-photon exchange effect of the neutron, e.g., the Friar and Zemach radius contributions, can be calculated based on neutron form factor parametrizations \cite{Tomalak:2018ysg}.
 
Also in the field of muonic atoms, the muX project\footref{section22}
determines nuclear charge radii of radioactive elements and rare
isotopes, e.g, $^{248}$Cm and $^{226}$Ra, through muonic X-ray
measurements. These are needed as input for atomic parity violation
experiments. In addition, muX probes nuclei that are at the end of a
double $\beta$ decay chain. These are interesting in view of possible
neutrinoless double $\beta$ decay that could occur if neutrinos were
Majorana particles. Two examples are the following $\beta^-\beta^-$
decays:
\begin{align}
\ce{&^{130}_{52}\text{Te}\stackrel{\beta^-}{\longrightarrow }{}^{130}_{53}\text{I}\stackrel{\beta^-}{\longrightarrow }{}^{130}_{54}\text{Xe}\nonumber ,\\
&^{82}_{34}\text{Se}\stackrel{\beta^-}{\longrightarrow }{}^{82}_{35}\text{Br}\stackrel{\beta^-}{\longrightarrow }{}^{82}_{36}\text{Kr}\nonumber {\begingroup\baselinestretch .\endgroup}}
\end{align}
Here one uses muon capture to study excited states of $^{130}$Xe and
$^{82}$Kr.  In the future, direct searches for BSM interactions
between muons and nuclei might be possible with the muX setup.

To further advance the precision of the few nucleon EFTs mentioned in
this section, the MuSun experiment\footref{section18} is studying
muon-capture on deuterium: $\mu^- d\rightarrow n n \nu_\mu$. The aim
is to determine the LEC of the axial-vector
four-nucleon interaction $d$ \cite{Pastore:2014iga}
\begin{equation}
\cL_{NN}=-2d(N^\dagger S\cdot u N)N^\dagger N,
\end{equation}
where $S^\mu$ is the nucleon covariant spin operator, $N(x)$ is the nucleon field, and $u_\mu$ is given below \eqref{LChPT}. Presently, this LEC has only been extracted from $A=3$ nuclei. The MuSun experiment has the potential for an improved extraction at the $20\,\%$ level.

\subsection{The free neutron}\label{th:secn}

In the previous section, we discussed nuclei and bound neutrons. In
the following, we discuss free neutrons provided by the Swiss
Spallation Neutron Source (SINQ) and the PSI Ultra Cold Neutron (UCN)
source~\cite{section04}. As we will see, the neutron experiments at
PSI are dedicated to BSM searches, and in particular, to the search
for CP violation in the light quark sector.

The neutron is unstable with a lifetime of about $880$\,s.
%$879.4 \pm0.6$ s %\cite{Zyla:2020zbs}. 
The long-standing tension between measurements with in-flight and stored neutrons has led to speculations that there could be `dark' BSM
decay channels~\cite{Fornal:2018eol, Czarnecki:2018okw}.  Within the
SM, the neutron decays into the proton, where the dominant decay
channel is the classical $\beta^-$ decay $n \rightarrow p
e^-\bar\nu_e$, described by the current-current interaction from the
Fermi theory, \eqref{th:L4F}. Besides the dominant $V-A$ structure of
the weak interaction, there could be small admixtures of scalar and
tensor couplings. Using the general formulation of Lee and Yang, which
is an older version of the parametrization in \eqref{th:matelJW}, the
$\beta^-$ decay reads \cite{Lee:1956qn}
\begin{align}
\langle p e^- \bar\nu_e\vert n\rangle=&\frac{G_F V_{ud}}{\sqrt{2}}\Bigg[\langle p\vert n\rangle \langle e^- \vert C_S-C_S' \gamma_5 \vert  \nu_e\rangle+\langle p\vert \gamma_\mu\vert n\rangle \langle e^- \vert \gamma^\mu\left(C_V-C_V' \gamma_5\right)\vert  \nu_e\rangle \nonumber \\
&+\nicefrac{1}{2}\,\langle p\vert \sigma_{\lambda\mu}\vert n\rangle \langle e^- \vert \sigma^{\lambda\mu}\left(C_T-C_T' \gamma_5\right)\vert  \nu_e\rangle-\langle p\vert \gamma_\mu\gamma_5\vert n\rangle \langle e^- \vert \gamma^\mu\gamma_5\left(C_A-C_A' \gamma_5\right)\vert  \nu_e\rangle\nonumber\\
&+\langle p\vert \gamma_5\vert n\rangle \langle e^- \vert \gamma_5\left(C_P-C_P' \gamma_5\right)\vert  \nu_e\rangle + \text{h.c.} \Bigg],\label{LeeYangbetadecay}
\end{align}
where $C_i^{(\prime)}$ are $10$ complex coupling constants. For the SM
with conserved vector current, $g_V=1$, the only non-vanishing
couplings are $C_V=C_V'=1$ and $C_A=C_A'=-g_A$. Parity violation is
assured if $C_i \neq 0$ and $C_i' \neq 0$. Time reversal violation
(TRV), or CP violation, is found if $\Im (C_i/C_j) \neq 0$ or $\Im
(C_i'/C_j )\neq 0$, i.e., if at least one coupling has an imaginary
phase relative to the others. The nTRV
experiment\hyperlink{ht:section15}{\footnote{\label{section15} Section 15: nTRV
  \cite{section15}.}} accessed the scalar and tensor couplings through
the measurement of the transverse polarization of electrons from the
decay of polarized free neutrons. At the present level of precision,
the results are in agreement with the SM, thus, setting constraints on
BSM physics. For a review on electroweak SM tests with nuclear $\beta$
decays see \cite{Severijns:2006dr}.

The observation of a nonzero permanent EDM of the neutron could be
interpreted as a signal of CP violating BSM interactions or a
measurement of the QCD $\theta$ parameter, see \eqref{thetaterm}. The
current best limit $|d_n| < 1.8\times 10^{-26}\,e$\,cm is 
from the nEDM experiment\hyperlink{ht:section27}{\footnote{\label{section27} Section 27: nEDM
  \cite{section27}.}} at PSI. This limit is still compatible with the
CKM-induced SM contributions to $d_n$, which are negligible as
explained below \eqref{th:F3df}. The n2EDM experiment will improve the sensitivity to $d_n$ by an order of magnitude and probe BSM physics at the multi-TeV scale~\cite{Engel:2013lsa}. The electric field of these experiments is of the order of $10^6\,$V/m. This is well below the critical electric field strength, $E_\mathrm{crit.} \sim 10^{23}$ V/m, that would be able to induce an EDM proportional to the neutron electric dipole
polarizability $d_\mathrm{ind.}=4\pi \alpha_{E1} \vec
E$~\cite{Hagelstein:2015egb}.  The nEDM spectrometer has also been used in indirect searches for Dark Matter (DM) candidates, e.g., mirror matter or axions and axion-like particles (ALPs).\hyperlink{ht:section28}{\footnote{\label{section28} Section 28: nEDMX \cite{section28}.}}

\subsection{The pion}\label{th:secpi}

Low-energy pion physics provides access to a large variety of
phenomena, ranging from strong non-perturbative dynamics over
electroweak precision tests to probes of BSM physics.  The pions are
stable in pure QCD and as asymptotic QCD states they play a special
role in many hadronic processes, where they appear as hadronic final
states. Pion interactions can be understood beyond the chiral expansion
by employing unitarity and analyticity of transition amplitudes, which
provide a means to resum pion-rescattering effects. Most notably,
$\pi\pi$ scattering has been accurately described in terms of the Roy
equations~\cite{Ananthanarayan:2000ht,Caprini:2011ky,GarciaMartin:2011cn},
and the resulting precise determination of the scattering phase shifts
provides a central input in the analysis of a host of other hadronic
processes at low energies.

An important probe of QCD at low energies is provided by the
interaction of pions with nucleons. Pionic atoms provide access to
$S$-wave $\pi N$ scattering lengths~\cite{Gasser:2007zt}, because the
strong interaction changes the spectrum compared to pure QED,
resulting in shifts of the energy levels and in finite widths of the
bound state. The most precise measurements of pionic hydrogen
and deuterium have been performed at PSI.\footref{section14} The $S$-wave
scattering lengths enter as important constraints in a dispersive
Roy--Steiner analysis of the $\pi N$ scattering
amplitude~\cite{Hoferichter:2015hva}.

Compared to pure strong dynamics in the isospin limit, both
electromagnetic effects and the mass difference between up and down
quarks generate small isospin-breaking corrections. The mass
difference of charged and neutral pions is understood to arise almost
exclusively from electromagnetic
effects~\cite{Das:1967it,Gasser:1983yg,Donoghue:1996zn}. This mass
difference $m_{\pi^-} - m_{\pi^0}$ has been determined with high
precision at PSI\hyperlink{ht:section12}{\footnote{\label{section12} Section 12: neutral pions
  \cite{section12}.}} starting from $(\pi^- p)$ bound states with
subsequent charge-exchange reaction $\pi^- p \to \pi^0 n$.  $m_{\pi^-}$ has
also been determined at PSI by measuring the
energy spectrum of pionic hydrogen $(\pi^- p)$.\hyperlink{ht:section10}{\footnote{\label{section10} Section 10: negative pions \cite{section10}.}}

In the presence of electromagnetism, the neutral pion is not a stable
particle, and decays predominantly into two photons. The decay results
from the anomalous non-conservation of the axial current that couples
to the pion. Quark-mass and electromagnetic corrections to the leading
Adler--Bell--Jackiw anomaly have been worked
out~\cite{Ananthanarayan:2002kj,Kampf:2009tk}. Further decay modes,
such as $\pi^0 \to e^+e^-\gamma$, $\pi^0\to4e$, and $\pi^0\to e^+e^-$
involve the transition $\pi^0\to\gamma^*\gamma^{(*)}$ with one or two
virtual photons. The transition form factor for this process has
received considerable interest in connection with hadronic
contributions to the muon anomalous magnetic
moment~\cite{Aoyama:2020ynm,Masjuan:2017tvw,Hoferichter:2018kwz,Gerardin:2019vio}.

Charged pions only decay due to the weak interaction. The hadronic part of the decay
rate for $\pi^+\to\ell^+\nu_\ell$ is governed by the pion decay constant $F_\pi$ of \eqref{th:pideccst}, whereas the leptonic part results in a helicity suppression by a factor $m_\ell^2$. Hence, the muonic decay mode dominates over the
electronic mode and has been used to
measure\hyperlink{ht:section11}{\footnote{\label{section11} Section 11: positive pions
  \cite{section11}.}} the mass of $\pi^+$. Several other decay modes
have been measured at PSI by the SINDRUM,\footref{section07}
PiBeta,\hyperlink{ht:section24}{\footnote{\label{section24} Section 24: PiBeta
  \cite{section24}.}} and PEN\hyperlink{ht:section25}{\footnote{\label{section25} Section 25: PEN
  \cite{section25}.}} experiments, including the radiative decays
$\pi^+\to\ell^+\nu_\ell\gamma$ and $\pi^+\to e^+\nu_e e^+e^-$ and pion
beta decay\footref{section24} $\pi^+\to\pi^0e^+\nu_e$. The theoretical description of the radiative decay $\pi^+\to\ell^+\nu_\ell\gamma$ is split into two parts, the so-called inner bremsstrahlung contributions (IB) and the structure-dependent terms (SD). The IB consist of the normal pion decay with additional emission of a photon from the charged external legs. This part depends on $F_\pi$. The SD terms require a more involved parametrization of the QCD effects in terms of two form factors. Apart from an axial form factor $F_A$ also a vector form factor $F_V$ contributes~\cite{Bryman:1982et}.  

The charged-pion decays probe the weak interaction in the low-energy
regime, where an excellent description is provided by Fermi's
effective theory of current-current interaction, or more generally the
LEFT framework explained in~\secref{th:secoverview}. The relevant
operator is
\begin{align}
	\cL_\mathrm{LEFT} \supset \sum_{i,j,k,l} C_{\substack{\nu edu \\ ijkl}}^{V,LL} (\bar\nu_i \gamma^\alpha P_L \ell_j)(\bar d_k \gamma_\alpha P_L u_l) + \mathrm{h.c.} ,
\end{align}
with flavor indices $i,j,k,l$ and the SM tree-level matching at the
weak scale given by \linebreak $C_{\substack{\nu edu \\ ijkl}}^{V,LL}
= -\frac{4G_F}{\sqrt{2}} \delta_{ij} V^\dagger_{kl}$. Therefore, the
pion decays probe the CKM matrix element $V_{ud}$, with a value of
$|V_{ud}| = 0.9739(27)$ resulting from the PiBeta measurement of pion
beta decay. Although precise, this value is not competitive with
determinations from superallowed nuclear beta
decays~\cite{Zyla:2020zbs}, which currently are in some tension with
first-row CKM unitarity. With the absence of nuclear structure aspects
and with radiative corrections under good theoretical
control~\cite{Cirigliano:2002ng}, pion beta decays are theoretically
clean but remain experimentally challenging due to the tiny branching
ratio $\sim 10^{-8}$.

Additional semileptonic operators in the LEFT Lagrangian with
different Dirac structures para\-metrize deviations from the SM and can
be probed by several pion decay modes~\cite{Cirigliano:2013xha}. E.g.,
strong constraints on the first-generation tensor-operator coefficient
$\Re(C_{\nu edu}^{T,RR})$ arise from the $\pi^+\to e^+\nu_e\gamma$
Dalitz-plot study of the PiBeta experiment.

\subsection{Conclusions}\label{th:concl}

Low-energy, high-precision experiments provide essential input to
improve our understanding of the fundamental interactions. They
complement and extend information obtained from the energy
frontier. EFTs are the theoretical tool of choice to describe and
interpret their results and indeed they are well suited to describe
both the SM and potential deviations therefrom in a model-independent
way. In particular it is possible, and crucial, to analyze if
potential deviations from the SM in different observables are linked
and have a common explanation. There are numerous examples where
low-energy constraints rule out apparently attractive new physics
scenarios. A broad and vigorous world-wide low-energy experimental
program is indispensable to make further progress in testing the SM
and searching for physics beyond. Past and future experiments at PSI
will continue to play their part in this challenge.


%%%%%%%%%% TODO: BIBNR Does the bibliography have more than 100 refs? Change 99 to 999.

%\bibliography{th}

\begin{thebibliography}{99}
%%%%%%%%%% END TODO: BIBNR

%%%%%%%%%% TODO: BBL IF BiBTeX was used: paste the contents of the .bbl file here
\providecommand{\url}[1]{\texttt{#1}}
\providecommand{\urlprefix}{URL }
\expandafter\ifx\csname urlstyle\endcsname\relax
  \providecommand{\doi}[1]{doi:\discretionary{}{}{}#1}\else
  \providecommand{\doi}{doi:\discretionary{}{}{}\begingroup
  \urlstyle{rm}\Url}\fi
\providecommand{\eprint}[2][]{\url{#2}}

\bibitem{section06} 
%%\textcolor{red}{TODO}
W. Fetscher, \textit{{Muon decay}}, SciPost Phys. Proc. \textbf{5}, 006 (2021), \doi{10.21468/SciPostPhysProc.5.006}.

%AUTHORS, \textit{TITLE}, JOURNAL \textbf{VOLUME}, PAGE/ARTICLE NUMBER (YEAR), \doi{DOI}.




\bibitem{section16} 
%%\textcolor{red}{TODO}
R. Carey, T. Gorringe and D. Hertzog, \textit{{Mulan: a part-per-million measurement of the muon lifetime and determination of the Fermi constant}}, SciPost Phys. Proc. \textbf{5}, 016 (2021), \doi{10.21468/SciPostPhysProc.5.016}.

%AUTHORS, \textit{TITLE}, JOURNAL \textbf{VOLUME}, PAGE/ARTICLE NUMBER (YEAR), \doi{DOI}.




\bibitem{section07} 
%%\textcolor{red}{TODO}
R. Eichler and C. Grab, \textit{{The SINDRUM-I experiment}}, SciPost Phys. Proc. \textbf{5}, 007 (2021), \doi{10.21468/SciPostPhysProc.5.007}.

%AUTHORS, \textit{TITLE}, JOURNAL \textbf{VOLUME}, PAGE/ARTICLE NUMBER (YEAR), \doi{DOI}.




\bibitem{section08} 
%%\textcolor{red}{TODO}
A. van der Schaaf, \textit{{Sindrum II }}, SciPost Phys. Proc. \textbf{5}, 008 (2021), \doi{10.21468/SciPostPhysProc.5.008}.

%AUTHORS, \textit{TITLE}, JOURNAL \textbf{VOLUME}, PAGE/ARTICLE NUMBER (YEAR), \doi{DOI}.




\bibitem{section19} 
%%\textcolor{red}{TODO}
A. Baldini and T. Mori, \textit{{MEG: Muon to electron and gamma}}, SciPost Phys. Proc. \textbf{5}, 019 (2021), \doi{10.21468/SciPostPhysProc.5.019}.

%AUTHORS, \textit{TITLE}, JOURNAL \textbf{VOLUME}, PAGE/ARTICLE NUMBER (YEAR), \doi{DOI}.




\bibitem{section20} 
%%\textcolor{red}{TODO}
F. Wauters, \emph{{The Mu3e experiment}}, SciPost Phys. Proc. \textbf{5}, 020 (2021), \doi{10.21468/SciPostPhysProc.5.020}.

%AUTHORS, \textit{TITLE}, JOURNAL \textbf{VOLUME}, PAGE/ARTICLE NUMBER (YEAR), \doi{DOI}.




\bibitem{section29} 
%%\textcolor{red}{TODO}
B. Ohayon, Z. Burkley and P. Crivelli, \textit{{Mspec: Muonium spectroscopy}}, SciPost Phys. Proc. \textbf{5}, 029 (2021), \doi{10.21468/SciPostPhysProc.5.029}.

%AUTHORS, \textit{TITLE}, JOURNAL \textbf{VOLUME}, PAGE/ARTICLE NUMBER (YEAR), \doi{DOI}.




\bibitem{section09} 
%%\textcolor{red}{TODO}
L. Willmann and K. Jungmann, \textit{{Muonium-antimuonium conversion}}, SciPost Phys. Proc. \textbf{5}, 009 (2021), \doi{10.21468/SciPostPhysProc.5.009}.

%AUTHORS, \textit{TITLE}, JOURNAL \textbf{VOLUME}, PAGE/ARTICLE NUMBER (YEAR), \doi{DOI}.




\bibitem{section21}
A. Antognini, F. Kottmann and R. Pohl, \textit{{Laser spectroscopy of light muonic atoms and the nuclear charge radii}}, SciPost Phys. Proc. \textbf{5}, 021 (2021), \doi{10.21468/SciPostPhysProc.5.021}.

%AUTHORS, \textit{TITLE}, JOURNAL \textbf{VOLUME}, PAGE/ARTICLE NUMBER (YEAR), \doi{DOI}.




\bibitem{section22} 
%%\textcolor{red}{TODO}
F. Wauters and A. Knecht, \textit{{The muX project}}, SciPost Phys. Proc. \textbf{5}, 020 (2021), \doi{10.21468/SciPostPhysProc.5.020}.

%AUTHORS, \textit{TITLE}, JOURNAL \textbf{VOLUME}, PAGE/ARTICLE NUMBER (YEAR), \doi{DOI}.




\bibitem{section23} 
%%\textcolor{red}{TODO}
E. Cline, J. Bernauer, E.J. Downie and R. Gilman, \textit{{MUSE: The MUon Scattering Experiment}}, SciPost Phys. Proc. \textbf{5}, 023 (2021), \doi{10.21468/SciPostPhysProc.5.023}.

%AUTHORS, \textit{TITLE}, JOURNAL \textbf{VOLUME}, PAGE/ARTICLE NUMBER (YEAR), \doi{DOI}.




\bibitem{section17} 
%%\textcolor{red}{TODO}
M. Hildebrandt and C. Petitjean, \textit{{MuCap: Muon capture on the proton}}, SciPost Phys. Proc. \textbf{5}, 017 (2021), \doi{10.21468/SciPostPhysProc.5.017}.

%AUTHORS, \textit{TITLE}, JOURNAL \textbf{VOLUME}, PAGE/ARTICLE NUMBER (YEAR), \doi{DOI}.




\bibitem{section18} 
%%\textcolor{red}{TODO}
P. Kammel, \textit{{MuSun - Muon capture on the deuteron}}, SciPost Phys. Proc. \textbf{5}, 018 (2021), \doi{10.21468/SciPostPhysProc.5.018}.

%AUTHORS, \textit{TITLE}, JOURNAL \textbf{VOLUME}, PAGE/ARTICLE NUMBER (YEAR), \doi{DOI}.




\bibitem{section14} 
%%\textcolor{red}{TODO}
D. Gotta and L. Simons, \textit{{Pionic hydrogen and deuterium}}, SciPost Phys. Proc. \textbf{5}, 014 (2021), \doi{10.21468/SciPostPhysProc.5.014}.

%AUTHORS, \textit{TITLE}, JOURNAL \textbf{VOLUME}, PAGE/ARTICLE NUMBER (YEAR), \doi{DOI}.




\bibitem{section26} 
%%\textcolor{red}{TODO}
M. Hori, H. Aghai-Khozani, A. S\'ot\'er, A. Dax and D. Barna, \textit{{Recent results of laser spectroscopy experiments of pionic helium atoms at PSI}}, SciPost Phys. Proc. \textbf{5}, 026 (2021), \doi{10.21468/SciPostPhysProc.5.026}.

%AUTHORS, \textit{TITLE}, JOURNAL \textbf{VOLUME}, PAGE/ARTICLE NUMBER (YEAR), \doi{DOI}.




\bibitem{section15} 
%%\textcolor{red}{TODO}
K. Bodek and A. Kozela, \textit{{Measurement of the transverse polarization of electrons emitted in neutron decay -- nTRV experiment}}, SciPost Phys. Proc. \textbf{5}, 015 (2021), \doi{10.21468/SciPostPhysProc.5.015}.

%AUTHORS, \textit{TITLE}, JOURNAL \textbf{VOLUME}, PAGE/ARTICLE NUMBER (YEAR), \doi{DOI}.




\bibitem{section27} 
%%\textcolor{red}{TODO}
G. Pignol and P. Schmidt-Wellenburg, \textit{{The search for the neutron electric dipole moment at PSI}}, SciPost Phys. Proc. \textbf{5}, 027 (2021), \doi{10.21468/SciPostPhysProc.5.027}.

%AUTHORS, \textit{TITLE}, JOURNAL \textbf{VOLUME}, PAGE/ARTICLE NUMBER (YEAR), \doi{DOI}.




\bibitem{section28} 
%%\textcolor{red}{TODO}
S. Roccia and G. Zsigmond, \textit{{Indirect searches for dark matter with the nEDM spectrometer}}, SciPost Phys. Proc. \textbf{5}, 028 (2021), \doi{10.21468/SciPostPhysProc.5.028}.

%AUTHORS, \textit{TITLE}, JOURNAL \textbf{VOLUME}, PAGE/ARTICLE NUMBER (YEAR), \doi{DOI}.




\bibitem{section10} 
%%\textcolor{red}{TODO}
M. Daum and D. Gotta, \textit{{The mass of the $\pi^-$ }}, SciPost Phys. Proc. \textbf{5}, 010 (2021), \doi{10.21468/SciPostPhysProc.5.010}.

%AUTHORS, \textit{TITLE}, JOURNAL \textbf{VOLUME}, PAGE/ARTICLE NUMBER (YEAR), \doi{DOI}.




\bibitem{section11} 
%%\textcolor{red}{TODO}
M. Daum and P.-R. Kettle, \textit{{The mass of the $\pi^+$ }}, SciPost Phys. Proc. \textbf{5}, 011 (2021), \doi{10.21468/SciPostPhysProc.5.011}.

%AUTHORS, \textit{TITLE}, JOURNAL \textbf{VOLUME}, PAGE/ARTICLE NUMBER (YEAR), \doi{DOI}.




\bibitem{section12} 
%%\textcolor{red}{TODO}
M. Daum and P.-R. Kettle, \textit{{The $\pi^0$ mass and the first experimental verification of Coulomb de-excitation in pinoic hydrogen}}, SciPost Phys. Proc. \textbf{5}, 012 (2021), \doi{10.21468/SciPostPhysProc.5.012}.

%AUTHORS, \textit{TITLE}, JOURNAL \textbf{VOLUME}, PAGE/ARTICLE NUMBER (YEAR), \doi{DOI}.




\bibitem{section24} 
%%\textcolor{red}{TODO}
D. Počanić, \textit{{The pion beta and radiative electronic decays}}, SciPost Phys. Proc. \textbf{5}, 024 (2021), \doi{10.21468/SciPostPhysProc.5.024}.

%AUTHORS, \textit{TITLE}, JOURNAL \textbf{VOLUME}, PAGE/ARTICLE NUMBER (YEAR), \doi{DOI}.




\bibitem{section25} 
%%\textcolor{red}{TODO}
D. Počanić, \textit{{Pion electronic decay and lepton universality}}, SciPost Phys. Proc. \textbf{5}, 025 (2021), \doi{10.21468/SciPostPhysProc.5.025}.

%AUTHORS, \textit{TITLE}, JOURNAL \textbf{VOLUME}, PAGE/ARTICLE NUMBER (YEAR), \doi{DOI}.




\bibitem{Jenkins:2017jig} 
%%\textcolor{red}{TODO}
%E. E. Jenkins, A. V. Manohar and P. Stoffer, \textit{{Low-Energy Effective Field Theory below the Electroweak Scale: Operators and Matching}}, JHEP \textbf{03}, 016 (2018), \doi{10.1007/JHEP03(2018)016}, \eprint{1709.04486}.

E. E. Jenkins, A. V. Manohar and P. Stoffer, \textit{Low-energy effective field theory below the electroweak scale: operators and matching}, J. High Energy Phys. \textbf{03}, 016 (2018), \doi{10.1007/JHEP03(2018)016}.




\bibitem{Peccei:1977hh} 
%%\textcolor{red}{TODO}
%R. D. Peccei and H. R. Quinn, \textit{{CP Conservation in the Presence of Instantons}}, Phys. Rev. Lett. \textbf{38}, 1440 (1977), \doi{10.1103/PhysRevLett.38.1440}.

R. D. Peccei and H. R. Quinn, \textit{CP Conservation in the presence of pseudoparticles}, Phys. Rev. Lett. \textbf{38}, 1440 (1977), \doi{10.1103/PhysRevLett.38.1440}.




\bibitem{Peccei:1977ur} 
%%\textcolor{red}{TODO}
%R. D. Peccei and H. R. Quinn, \textit{{Constraints Imposed by CP Conservation in the Presence of Instantons}}, Phys. Rev. D \textbf{16}, 1791 (1977), \doi{10.1103/PhysRevD.16.1791}.

R. D. Peccei and H. R. Quinn, \textit{Constraints imposed by CP conservation in the presence of pseudoparticles}, Phys. Rev. D \textbf{16}, 1791 (1977), \doi{10.1103/PhysRevD.16.1791}.




\bibitem{Wilczek:1977md} 
%%\textcolor{red}{TODO}
%F. Wilczek and A. Zee, \textit{{Instantons and Spin Forces Between Massive Quarks}}, Phys. Rev. Lett. \textbf{40}, 83 (1978), \doi{10.1103/PhysRevLett.40.83}.

F. Wilczek and A. Zee, \textit{Instantons and spin forces between massive quarks}, Phys. Rev. Lett. \textbf{40}, 83 (1978), \doi{10.1103/PhysRevLett.40.83}.




\bibitem{Weinberg:1977ma} 
%%\textcolor{red}{TODO}
%S. Weinberg, \textit{{A New Light Boson?}}, Phys. Rev. Lett. \textbf{40}, 223 (1978), \doi{10.1103/PhysRevLett.40.223}.

S. Weinberg, \textit{A new light boson?}, Phys. Rev. Lett. \textbf{40}, 223 (1978), \doi{10.1103/PhysRevLett.40.223}.




\bibitem{Michel:1949qe} 
%%\textcolor{red}{TODO}
%L. Michel, \textit{{Interaction between four half spin particles and the decay of the $\mu$ meson}}, Proc. Phys. Soc. \textbf{A63}, 514 (1950), \doi{10.1088/0370-1298/63/5/311}, [,45(1949)].

L. Michel, \textit{Interaction between four half-spin particles and the decay of the  $\mu$-meson}, Proc. Phys. Soc. A \textbf{63}, 514 (1950), \doi{10.1088/0370-1298/63/5/311}.




\bibitem{Fetscher:1986uj} 
%%\textcolor{red}{TODO}
%W. Fetscher, H. J. Gerber and K. F. Johnson, \textit{{Muon Decay: Complete Determination of the Interaction and Comparison with the Standard Model}}, Phys. Lett. \textbf{B173}, 102 (1986), \doi{10.1016/0370-2693(86)91239-6}.

W. Fetscher, H.-J. Gerber and K. F. Johnson, \textit{Muon decay: complete determination of the interaction and comparison with the standard model}, Phys. Lett. B \textbf{173}, 102 (1986), \doi{10.1016/0370-2693(86)91239-6}.




\bibitem{Kuno:1999jp} 
%%\textcolor{red}{TODO}
%Y. Kuno and Y. Okada, \textit{{Muon decay and physics beyond the standard model}}, Rev. Mod. Phys. \textbf{73}, 151 (2001), \doi{10.1103/RevModPhys.73.151}, \eprint{hep-ph/9909265}.

Y. Kuno and Y. Okada, \textit{Muon decay and physics beyond the standard model}, Rev. Mod. Phys. \textbf{73}, 151 (2001), \doi{10.1103/RevModPhys.73.151}.




\bibitem{Weinberg:1979sa} 
%\textcolor{red}{TODO}
%S. Weinberg, \textit{{Baryon and Lepton Nonconserving Processes}}, Phys. Rev. Lett. \textbf{43}, 1566 (1979), \doi{10.1103/PhysRevLett.43.1566}.

S. Weinberg, \textit{Baryon- and lepton-nonconserving processes}, Phys. Rev. Lett. \textbf{43}, 1566 (1979), \doi{10.1103/PhysRevLett.43.1566}.




\bibitem{Grzadkowski:2010es} 
%\textcolor{red}{TODO}
%B. Grzadkowski, M. Iskrzynski, M. Misiak and J. Rosiek, \textit{{Dimension-Six Terms in the Standard Model Lagrangian}}, JHEP \textbf{1010}, 085 (2010), \doi{10.1007/JHEP10(2010)085}, \eprint{1008.4884}.

B. Grzadkowski, M. Iskrzyński, M. Misiak and J. Rosiek, \textit{Dimension-six terms in the Standard Model Lagrangian}, J. High Energy Phys. \textbf{10}, 085 (2010), \doi{10.1007/JHEP10(2010)085}.




\bibitem{Jenkins:2013zja} 
%\textcolor{red}{TODO}
%E. E. Jenkins, A. V. Manohar and M. Trott, \textit{{Renormalization Group Evolution of the Standard Model Dimension Six Operators I: Formalism and lambda Dependence}}, JHEP \textbf{1310}, 087 (2013), \doi{10.1007/JHEP10(2013)087}, \eprint{1308.2627}.

E. E. Jenkins, A. V. Manohar and M. Trott, \textit{Renormalization group evolution of the standard model dimension six operators. I: formalism and $\lambda$ dependence}, J. High Energy Phys. \textbf{10}, 087 (2013), \doi{10.1007/JHEP10(2013)087}.




\bibitem{Jenkins:2013wua} 
%\textcolor{red}{TODO}
%E. E. Jenkins, A. V. Manohar and M. Trott, \textit{{Renormalization Group Evolution of the Standard Model Dimension Six Operators II: Yukawa Dependence}}, JHEP \textbf{1401}, 035 (2014), \doi{10.1007/JHEP01(2014)035}, \eprint{1310.4838}.

E. E. Jenkins, A. V. Manohar and M. Trott, \textit{Renormalization group evolution of the Standard Model dimension six operators II: Yukawa dependence}, J. High Energy Phys. \textbf{01}, 035 (2014), \doi{10.1007/JHEP01(2014)035}.




\bibitem{Alonso:2013hga} 
%\textcolor{red}{TODO}
%R. Alonso, E. E. Jenkins, A. V. Manohar and M. Trott, \textit{{Renormalization Group Evolution of the Standard Model Dimension Six Operators III: Gauge Coupling Dependence and Phenomenology}}, JHEP \textbf{1404}, 159 (2014), \doi{10.1007/JHEP04(2014)159}, \eprint{1312.2014}.

R. Alonso, E. E. Jenkins, A. V. Manohar and M. Trott, \textit{Renormalization group evolution of the Standard Model dimension six operators III: gauge coupling dependence and phenomenology}, J. High Energy Phys. \textbf{04}, 159 (2014), \doi{10.1007/JHEP04(2014)159}.




\bibitem{Jenkins:2017dyc} 
%\textcolor{red}{TODO}
%E. E. Jenkins, A. V. Manohar and P. Stoffer, \textit{{Low-Energy Effective Field Theory below the Electroweak Scale: Anomalous Dimensions}}, JHEP \textbf{01}, 084 (2018), \doi{10.1007/JHEP01(2018)084}, \eprint{1711.05270}.

E. E. Jenkins, A. V. Manohar and P. Stoffer, \textit{Low-energy effective field theory below the electroweak scale: anomalous dimensions}, J. High Energy Phys. \textbf{01}, 084 (2018), \doi{10.1007/JHEP01(2018)084}.




\bibitem{Dekens:2019ept} 
%\textcolor{red}{TODO}
%W. Dekens and P. Stoffer, \textit{{Low-energy effective field theory below the electroweak scale: matching at one loop}}, JHEP \textbf{10}, 197 (2019), \doi{10.1007/JHEP10(2019)197}, \eprint{1908.05295}.

W. Dekens and P. Stoffer, \textit{Low-energy effective field theory below the electroweak scale: matching at one loop}, J. High Energy Phys. \textbf{10}, 197 (2019), \doi{10.1007/JHEP10(2019)197}.




\bibitem{Czarnecki:1997bu} 
%\textcolor{red}{TODO}
%A. Czarnecki and B. Krause, \textit{{Neutron electric dipole moment in the standard model: Valence quark contributions}}, Phys. Rev. Lett. \textbf{78}, 4339 (1997), \doi{10.1103/PhysRevLett.78.4339}, \eprint{hep-ph/9704355}.

A. Czarnecki and B. Krause, \textit{Neutron electric dipole moment in the Standard Model: Complete three-loop calculation of the valence quark contributions}, Phys. Rev. Lett. \textbf{78}, 4339 (1997), \doi{10.1103/PhysRevLett.78.4339}.




\bibitem{Pospelov:1991zt} 
%\textcolor{red}{TODO}
M. E. Pospelov and I. B. Khriplovich, \textit{{Electric dipole moment of the W boson and the electron in the Kobayashi-Maskawa model}}, Sov. J. Nucl. Phys. \textbf{53}, 638 (1991).

%AUTHORS, \textit{TITLE}, JOURNAL \textbf{VOLUME}, PAGE/ARTICLE NUMBER (YEAR), \doi{DOI}.




\bibitem{Pospelov:1999ha} 
%\textcolor{red}{TODO}
%M. Pospelov and A. Ritz, \textit{{Theta induced electric dipole moment of the neutron via QCD sum rules}}, Phys. Rev. Lett. \textbf{83}, 2526 (1999), \doi{10.1103/PhysRevLett.83.2526}, \eprint{hep-ph/9904483}.

M. Pospelov and A. Ritz, \textit{Theta-induced electric dipole moment of the neutron via QCD sum rules}, Phys. Rev. Lett. \textbf{83}, 2526 (1999), \doi{10.1103/PhysRevLett.83.2526}.




\bibitem{Pospelov:2005pr} 
%\textcolor{red}{TODO}
%M. Pospelov and A. Ritz, \textit{{Electric dipole moments as probes of new physics}}, Annals Phys. \textbf{318}, 119 (2005), \doi{10.1016/j.aop.2005.04.002}, \eprint{hep-ph/0504231}.

M. Pospelov and A. Ritz, \textit{Electric dipole moments as probes of new physics}, Ann. Phys. \textbf{318}, 119 (2005), \doi{10.1016/j.aop.2005.04.002}.




\bibitem{Engel:2013lsa}
 %\textcolor{red}{TODO}
%J. Engel, M. J. Ramsey-Musolf and U. van Kolck, \textit{{Electric Dipole Moments of Nucleons, Nuclei, and Atoms: The Standard Model and Beyond}}, Prog. Part. Nucl. Phys. \textbf{71}, 21 (2013), \doi{10.1016/j.ppnp.2013.03.003}, \eprint{1303.2371}.

J. Engel, M. J. Ramsey-Musolf and U. van Kolck, \textit{Electric dipole moments of nucleons, nuclei, and atoms: The Standard Model and beyond}, Prog. Part. Nucl. Phys. \textbf{71}, 21 (2013), \doi{10.1016/j.ppnp.2013.03.003}.




\bibitem{Musolf:1990sa} 
%\textcolor{red}{TODO}
%M. J. Musolf and B. R. Holstein, \textit{{Observability of the anapole moment and neutrino charge radius}}, Phys. Rev. D \textbf{43}, 2956 (1991), \doi{10.1103/PhysRevD.43.2956}.

M. J. Musolf and B. R. Holstein, \textit{Observability of the anapole moment and neutrino charge radius}, Phys. Rev. D \textbf{43}, 2956 (1991), \doi{10.1103/PhysRevD.43.2956}.




\bibitem{Weinberg:1978kz} 
%\textcolor{red}{TODO}
%S. Weinberg, \textit{{Phenomenological Lagrangians}}, Physica A \textbf{96}(1-2), 327 (1979), \doi{10.1016/0378-4371(79)90223-1}.

S. Weinberg, \textit{Phenomenological Lagrangians}, Phys. A: Stat. Mech. Appl. \textbf{96}, 327 (1979), \doi{10.1016/0378-4371(79)90223-1}.




\bibitem{Gasser:1983yg} 
%\textcolor{red}{TODO}
%J. Gasser and H. Leutwyler, \textit{{Chiral Perturbation Theory to One Loop}}, Annals Phys. \textbf{158}, 142 (1984), \doi{10.1016/0003-4916(84)90242-2}.

J. Gasser and H. Leutwyler, \textit{Chiral perturbation theory to one loop}, Ann. Phys. \textbf{158}, 142 (1984), \doi{10.1016/0003-4916(84)90242-2}.




\bibitem{Leutwyler:1993iq} 
%\textcolor{red}{TODO}
%H. Leutwyler, \textit{{On the foundations of chiral perturbation theory}}, Annals Phys. \textbf{235}, 165 (1994), \doi{10.1006/aphy.1994.1094}, \eprint{hep-ph/9311274}.

H. Leutwyler, \textit{On the foundations of chiral perturbation theory}, Ann. Phys. \textbf{235}, 165 (1994), \doi{10.1006/aphy.1994.1094}.




\bibitem{Gell-Mann:1968hlm} 
%\textcolor{red}{TODO}
%M. Gell-Mann, R. J. Oakes and B. Renner, \textit{{Behavior of current divergences under SU(3) x SU(3)}}, Phys. Rev. \textbf{175}, 2195 (1968), \doi{10.1103/PhysRev.175.2195}.

M. Gell-Mann, R. J. Oakes and B. Renner, \textit{Behavior of current divergences under $SU_3\times SU_3$}, Phys. Rev. \textbf{175}, 2195 (1968), \doi{10.1103/PhysRev.175.2195}.




\bibitem{Fearing:1994ga} 
%\textcolor{red}{TODO}
%H.~W. Fearing and S.~Scherer, \textit{{Extension of the chiral perturbation theory meson Lagrangian to order p(6)}}, Phys. Rev. D \textbf{53}, 315 (1996), \doi{10.1103/PhysRevD.53.315}, \eprint{hep-ph/9408346}.

H. W. Fearing and S. Scherer, \textit{Extension of the chiral perturbation theory meson Lagrangian to order ${\mathit{p}}^{6}$}, Phys. Rev. D \textbf{53}, 315 (1996), \doi{10.1103/PhysRevD.53.315}.




\bibitem{Bijnens:1999hw} 
%\textcolor{red}{TODO}
%J.~Bijnens, G.~Colangelo and G.~Ecker, \textit{{Renormalization of chiral perturbation theory to order p**6}}, Annals Phys. \textbf{280}, 100 (2000), \doi{10.1006/aphy.1999.5982}, \eprint{hep-ph/9907333}.

J. Bijnens, G. Colangelo and G. Ecker, \textit{Renormalization of chiral perturbation theory to order $p^6$}, Ann. Phys. \textbf{280}, 100 (2000), \doi{10.1006/aphy.1999.5982}.




\bibitem{Bijnens:1999sh} 
%\textcolor{red}{TODO}
%J.~Bijnens, G.~Colangelo and G.~Ecker, \textit{{The Mesonic chiral Lagrangian of order p**6}}, JHEP \textbf{02}, 020 (1999), \doi{10.1088/1126-6708/1999/02/020}, \eprint{hep-ph/9902437}.

J. Bijnens, G. Colangelo and G. Ecker, \textit{The mesonic chiral lagrangean of order $p^6$}, J. High Energy Phys. \textbf{02}, 020 (1999), \doi{10.1088/1126-6708/1999/02/020}.




\bibitem{Gasser:1984gg} 
%\textcolor{red}{TODO}
%J.~Gasser and H.~Leutwyler, \textit{{Chiral Perturbation Theory: Expansions in the Mass of the Strange Quark}}, Nucl. Phys. B \textbf{250}, 465 (1985), \doi{10.1016/0550-3213(85)90492-4}.

J. Gasser and H. Leutwyler, \textit{Chiral perturbation theory: Expansions in the mass of the strange quark}, Nucl. Phys. B \textbf{250}, 465 (1985), \doi{10.1016/0550-3213(85)90492-4}.




\bibitem{Gasser:1987rb} 
%\textcolor{red}{TODO}
%J.~Gasser, M.~E. Sainio and A.~Svarc, \textit{{Nucleons with Chiral Loops}}, Nucl. Phys. B \textbf{307}, 779 (1988), \doi{10.1016/0550-3213(88)90108-3}.

J. Gasser, M. E. Sainio and A. Švarc, \textit{Nucleons with chiral loops}, Nucl. Phys. B \textbf{307}, 779 (1988), \doi{10.1016/0550-3213(88)90108-3}.




\bibitem{Jenkins:1990jv} 
%\textcolor{red}{TODO}
%E.~E. Jenkins and A.~V. Manohar, \textit{{Baryon chiral perturbation theory using a heavy fermion Lagrangian}}, Phys. Lett. B \textbf{255}, 558 (1991), \doi{10.1016/0370-2693(91)90266-S}.

E. Jenkins and A. V. Manohar, \textit{Baryon chiral perturbation theory using a heavy fermion lagrangian}, Phys. Lett. B \textbf{255}, 558 (1991), \doi{10.1016/0370-2693(91)90266-S}.




\bibitem{Bernard:1992qa}
 %\textcolor{red}{TODO}
%V.~Bernard, N.~Kaiser, J.~Kambor and U.~G. Meissner, \textit{{Chiral structure of the nucleon}}, Nucl. Phys. B \textbf{388}, 315 (1992), \doi{10.1016/0550-3213(92)90615-I}.

V. Bernard, N. Kaiser, J. Kambor and U.-G. Meißner, \textit{Chiral structure of the nucleon}, Nucl. Phys. B \textbf{388}, 315 (1992), \doi{10.1016/0550-3213(92)90615-I}.




\bibitem{Becher:1999he} 
%\textcolor{red}{TODO}
%T.~Becher and H.~Leutwyler, \textit{{Baryon chiral perturbation theory in manifestly Lorentz invariant form}}, Eur. Phys. J. C \textbf{9}, 643 (1999), \doi{10.1007/PL00021673}, \eprint{hep-ph/9901384}.

T. Becher and H. Leutwyler, \textit{Baryon chiral perturbation theory in manifestly Lorentz invariant form}, Eur. Phys. J. C \textbf{9}, 643 (1999), \doi{10.1007/PL00021673}.




\bibitem{Goldberger:1958zz} 
%\textcolor{red}{TODO}
%M.~L. Goldberger and S.~B. Treiman, \textit{{Conserved Currents in the Theory of Fermi Interactions}}, Phys. Rev. \textbf{110}, 1478 (1958), \doi{10.1103/PhysRev.110.1478}.

M. L. Goldberger and S. B. Treiman, \textit{Conserved currents in the theory of Fermi interactions}, Phys. Rev. \textbf{110}, 1478 (1958), \doi{10.1103/PhysRev.110.1478}.




\bibitem{Berman:196220}
 %\textcolor{red}{TODO}
%S.~Berman and A.~Sirlin, \textit{Some considerations on the radiative corrections to muon and neutron decay}, Annals of Physics \textbf{20}(1), 20 (1962), \doi{http://dx.doi.org/10.1016/0003-4916(62)90114-8}.

S. M. Berman and A. Sirlin, \textit{Some considerations on the radiative corrections to muon and neutron decay}, Ann. Phys. \textbf{20}, 20 (1962), \doi{10.1016/0003-4916(62)90114-8}.




\bibitem{vanRitbergen:1998yd} 
%\textcolor{red}{TODO}
%T.~van Ritbergen and R.~G. Stuart, \textit{{Complete two loop quantum electrodynamic contributions to the muon lifetime in the Fermi model}}, Phys.Rev.Lett. \textbf{82}, 488 (1999), \doi{10.1103/PhysRevLett.82.488}, \eprint{hep-ph/9808283}.

T. van Ritbergen and R. G. Stuart, \textit{Complete 2-loop quantum electrodynamic contributions to the muon lifetime in the Fermi model}, Phys. Rev. Lett. \textbf{82}, 488 (1999), \doi{10.1103/PhysRevLett.82.488}.




\bibitem{Anastasiou:2005pn}
 %\textcolor{red}{TODO}
%C.~Anastasiou, K.~Melnikov and F.~Petriello, \textit{{The Electron energy spectrum in muon decay through O(alpha**2)}}, JHEP \textbf{0709}, 014 (2007), \doi{10.1088/1126-6708/2007/09/014}, \eprint{hep-ph/0505069}.

C. Anastasiou, K. Melnikov and F. Petriello, \textit{The electron energy spectrum in muon decay through $\mathcal{O}(\alpha^2)$}, J. High Energy Phys. \textbf{09}, 014 (2007), \doi{10.1088/1126-6708/2007/09/014}.




\bibitem{Pak:2008qt} 
%\textcolor{red}{TODO}
%A.~Pak and A.~Czarnecki, \textit{{Mass effects in muon and semileptonic $b\to c$ decays}}, Phys. Rev. Lett. \textbf{100}, 241807 (2008), \doi{10.1103/PhysRevLett.100.241807}, \eprint{0803.0960}.

A. Pak and A. Czarnecki, \textit{Mass effects in muon and semileptonic $b\rightarrow c$ decays}, Phys. Rev. Lett. \textbf{100}, 241807 (2008), \doi{10.1103/PhysRevLett.100.241807}.




\bibitem{Engel:2019nfw} 
%\textcolor{red}{TODO}
%T.~Engel, A.~Signer and Y.~Ulrich, \textit{{A subtraction scheme for massive QED}}, JHEP \textbf{01}, 085 (2020), \doi{10.1007/JHEP01(2020)085}, [JHEP20,085(2020)], \eprint{1909.10244}.

T. Engel, A. Signer and Y. Ulrich, \textit{A subtraction scheme for massive QED}, J. High Energy Phys. \textbf{01}, 085 (2020), \doi{10.1007/JHEP01(2020)085}.




\bibitem{Fael:2015gua} 
%\textcolor{red}{TODO}
%M.~Fael, L.~Mercolli and M.~Passera, \textit{{Radiative $\mu$ and $\tau$ leptonic decays at NLO}}, JHEP \textbf{07}, 153 (2015), \doi{10.1007/JHEP07(2015)153}, \eprint{1506.03416}.

M. Fael, L. Mercolli and M. Passera, \textit{Radiative $\mu$ and $\tau$ leptonic decays at NLO}, J. High Energy Phys. \textbf{07}, 153 (2015), \doi{10.1007/JHEP07(2015)153}.




\bibitem{Pruna:2017upz} 
%\textcolor{red}{TODO}
%G.~M. Pruna, A.~Signer and Y.~Ulrich, \textit{{Fully differential NLO predictions for the radiative decay of muons and taus}}, Phys. Lett. \textbf{B772}, 452 (2017), \doi{10.1016/j.physletb.2017.07.008}, \eprint{1705.03782}.

G. M. Pruna, A. Signer and Y. Ulrich, \textit{Fully differential NLO predictions for the radiative decay of muons and taus}, Phys. Lett. B \textbf{772}, 452 (2017), \doi{10.1016/j.physletb.2017.07.008}.




\bibitem{Pruna:2016spf} 
%\textcolor{red}{TODO}
%G.~M. Pruna, A.~Signer and Y.~Ulrich, \textit{{Fully differential NLO predictions for the rare muon decay}}, Phys. Lett. \textbf{B765}, 280 (2017), \doi{10.1016/j.physletb.2016.12.039}, \eprint{1611.03617}.

G. M. Pruna, A. Signer and Y. Ulrich, \textit{Fully differential NLO predictions for the rare muon decay}, Phys. Lett. B \textbf{765}, 280 (2017), \doi{10.1016/j.physletb.2016.12.039}.




\bibitem{Fael:2016yle} 
%\textcolor{red}{TODO}
%M.~Fael and C.~Greub, \textit{{Next-to-leading order prediction for the decay $ \mu \to e (ee)\nu\bar{\nu}$}}, JHEP \textbf{01}, 084 (2017), \doi{10.1007/JHEP01(2017)084}, \eprint{1611.03726}.

M. Fael and C. Greub, \textit{Next-to-leading order prediction for the decay $ \mu \rightarrow e (e^+e^-)\nu\bar{\nu}$}, J. High Energy Phys. \textbf{01}, 084 (2017), \doi{10.1007/JHEP01(2017)084}.




\bibitem{Adam:2013gfn} 
%\textcolor{red}{TODO}

A. M. Baldini \textit{et al.}, \textit{{Measurement of the radiative decay of polarized muons in the MEG experiment}}, Eur. Phys. J. C \textbf{76}, 108 (2016), \doi{10.1140/epjc/s10052-016-3947-6}.

%None, \textit{Measurement of the radiative decay of polarized muons in the MEG experiment}, Eur. Phys. J. C \textbf{76}, 108 (2016), \doi{10.1140/epjc/s10052-016-3947-6}.




\bibitem{Pocanic:2014mya} 
%\textcolor{red}{TODO}

D. Počanić \textit{et al.}, \textit{{New results in rare allowed muon and pion decays}}, Int. J. Mod. Phys. Conf. Ser. \textbf{35}, 1460437 (2014), \doi{10.1142/S2010194514604372}.

%D. POČANIĆ \textit{et al.}, \textit{NEW RESULTS IN RARE ALLOWED MUON AND PION DECAYS}, Int. J. Mod. Phys. Conf. Ser. \textbf{35}, 1460437 (2014), \doi{10.1142/S2010194514604372}.




\bibitem{Petcov:1976ff} 
%\textcolor{red}{TODO}

S. T. Petcov, \textit{{The Processes $\mu \rightarrow e + \gamma, \mu \rightarrow e + \overline{e}, \nu' \rightarrow \nu + \gamma$ in the Weinberg-Salam model with neutrino mixing}}, Sov. J. Nucl. Phys. \textbf{25}, 340 (1977), [Erratum: Sov. J. Nucl. Phys. \textbf{25}, 698 (1977), Erratum: Yad. Fiz. \textbf{25}, 1336 (1977)].

%AUTHORS, \textit{TITLE}, JOURNAL \textbf{VOLUME}, PAGE/ARTICLE NUMBER (YEAR), \doi{DOI}.




\bibitem{Crivellin:2017rmk} 
%\textcolor{red}{TODO}
%A.~Crivellin, S.~Davidson, G.~M. Pruna and A.~Signer, \textit{{Renormalisation-group improved analysis of $\mu\to e$ processes in a systematic effective-field-theory approach}}, JHEP \textbf{05}, 117 (2017), \doi{10.1007/JHEP05(2017)117}, \eprint{1702.03020}.

A. Crivellin, S. Davidson, G. M. Pruna and A. Signer, \textit{Renormalisation-group improved analysis of $\mu\to e$ processes in a systematic effective-field-theory approach}, J. High Energy Phys. \textbf{05}, 117 (2017), \doi{10.1007/JHEP05(2017)117}.




\bibitem{Dekens:2018pbu} 
%\textcolor{red}{TODO}
%W.~Dekens, E.~E. Jenkins, A.~V. Manohar and P.~Stoffer, \textit{{Non-perturbative effects in $\mu \to e \gamma$}}, JHEP \textbf{01}, 088 (2019), \doi{10.1007/JHEP01(2019)088}, \eprint{1810.05675}.

W. Dekens, E. E. Jenkins, A. V. Manohar and P. Stoffer, \textit{Non-perturbative effects in $\mu \to e \gamma$}, J. High Energy Phys. \textbf{01}, 088 (2019), \doi{10.1007/JHEP01(2019)088}.




\bibitem{Kitano:2002mt} 
%\textcolor{red}{TODO}

R. Kitano, M. Koike and Y. Okada, \textit{{Detailed calculation of lepton flavor violating muon-electron conversion rate for various nuclei}}, Phys. Rev. D \textbf{66}, 096002 (2002), \doi{10.1103/PhysRevD.76.059902}, [Erratum: Phys. Rev. D \textbf{76}, 059902 (2007), \doi{10.1103/PhysRevD.66.096002}].

%AUTHORS, \textit{TITLE}, JOURNAL \textbf{VOLUME}, PAGE/ARTICLE NUMBER (YEAR), \doi{DOI}.




\bibitem{Cirigliano:2009bz} 
%\textcolor{red}{TODO}
%V.~Cirigliano, R.~Kitano, Y.~Okada and P.~Tuzon, \textit{{On the model discriminating power of $\mu \to e$ conversion in nuclei}}, Phys. Rev. \textbf{D80}, 013002 (2009), \doi{10.1103/PhysRevD.80.013002}, \eprint{0904.0957}.

V. Cirigliano, R. Kitano, Y. Okada and P. Tuzon, \textit{Model discriminating power of $\mu \to e$ conversion in nuclei}, Phys. Rev. D \textbf{80}, 013002 (2009), \doi{10.1103/PhysRevD.80.013002}.




\bibitem{Czarnecki:2014cxa} 
%\textcolor{red}{TODO}
%A.~Czarnecki, M.~Dowling, X.~Garcia~i Tormo, W.~J. Marciano and R.~Szafron, \textit{{Michel decay spectrum for a muon bound to a nucleus}}, Phys. Rev. \textbf{D90}(9), 093002 (2014), \doi{10.1103/PhysRevD.90.093002}, \eprint{1406.3575}.

A. Czarnecki, M. Dowling, X. Garcia i Tormo, W. J. Marciano and R. Szafron, \textit{Michel decay spectrum for a muon bound to a nucleus}, Phys. Rev. D \textbf{90}, 093002 (2014), \doi{10.1103/PhysRevD.90.093002}.




\bibitem{Antognini:2018nhb} 
%\textcolor{red}{TODO}
%A.~Antognini, D.~M. Kaplan, K.~Kirch, A.~Knecht, D.~C. Mancini, J.~D. Phillips, T.~J. Phillips, R.~D. Reasenberg, T.~J. Roberts and A.~Soter, \textit{{Studying Antimatter Gravity with Muonium}}, Atoms \textbf{6}(2), 17 (2018), \doi{10.3390/atoms6020017}, \eprint{1802.01438}.

A. Antognini \textit{et al.}, \textit{Studying antimatter gravity with muonium}, Atoms \textbf{6}, 17 (2018), \doi{10.3390/atoms6020017}.




\bibitem{Aoyama:2020ynm} 
%\textcolor{red}{TODO}
%T.~Aoyama \textit{et~al.}, \textit{{The anomalous magnetic moment of the muon in the Standard Model}}, Phys. Rept. \textbf{887}, 1 (2020), \doi{10.1016/j.physrep.2020.07.006}, \eprint{2006.04822}.

T. Aoyama \textit{et al.}, \textit{The anomalous magnetic moment of the muon in the Standard Model}, Phys. Rep. \textbf{887}, 1 (2020), \doi{10.1016/j.physrep.2020.07.006}.




\bibitem{Kirch:2020lbo} 
%\textcolor{red}{TODO}
%K.~Kirch and P.~Schmidt-Wellenburg, \textit{{Search for electric dipole moments}}, EPJ Web Conf. \textbf{234}, 01007 (2020), \doi{10.1051/epjconf/202023401007}, \eprint{2003.00717}.

K. Kirch and P. Schmidt-Wellenburg, \textit{Search for electric dipole moments}, EPJ Web Conf. \textbf{234}, 01007 (2020), \doi{10.1051/epjconf/202023401007}.




\bibitem{Perdrisat:2006hj} 
%\textcolor{red}{TODO}
%C.~F. Perdrisat, V.~Punjabi and M.~Vanderhaeghen, \textit{{Nucleon Electromagnetic Form Factors}}, Prog. Part. Nucl. Phys. \textbf{59}, 694 (2007), \doi{10.1016/j.ppnp.2007.05.001}, \eprint{hep-ph/0612014}.

C. F. Perdrisat, V. Punjabi and M. Vanderhaeghen, \textit{Nucleon electromagnetic form factors}, Prog. Part. Nucl. Phys. \textbf{59}, 694 (2007), \doi{10.1016/j.ppnp.2007.05.001}.




\bibitem{Pohl:2013yb}
 %\textcolor{red}{TODO}
%R.~Pohl, R.~Gilman, G.~A. Miller and K.~Pachucki, \textit{{Muonic hydrogen and the proton radius puzzle}}, Ann. Rev. Nucl. Part. Sci. \textbf{63}, 175 (2013), \doi{10.1146/annurev-nucl-102212-170627}, \eprint{1301.0905}.

R. Pohl, R. Gilman, G. A. Miller and K. Pachucki, \textit{Muonic hydrogen and the proton radius puzzle}, Annu. Rev. Nucl. Part. Sci. \textbf{63}, 175 (2013), \doi{10.1146/annurev-nucl-102212-170627}.




\bibitem{Carlson:2015jba} 
%\textcolor{red}{TODO}
%⎄C.~E. Carlson, \textit{{The Proton Radius Puzzle}}, Prog. Part. Nucl. Phys. \textbf{82}, 59 (2015), \doi{10.1016/j.ppnp.2015.01.002}, \eprint{1502.05314}.

C. E. Carlson, \textit{The proton radius puzzle}, Prog. Part. Nucl. Phys. \textbf{82}, 59 (2015), \doi{10.1016/j.ppnp.2015.01.002}.




\bibitem{Hammer:2019uab} 
%\textcolor{red}{TODO}
%H.-W. Hammer and U.-G. Mei\ss{}ner, \textit{{The proton radius: From a puzzle to precision}}, Sci. Bull. \textbf{65}, 257 (2020), \doi{10.1016/j.scib.2019.12.012}, \eprint{1912.03881}.

H.-W. Hammer and U.-G. Meißner, \textit{The proton radius: from a puzzle to precision}, Sci. Bull. \textbf{65}, 257 (2020), \doi{10.1016/j.scib.2019.12.012}.




\bibitem{Antognini:2013jkc} 
%\textcolor{red}{TODO}
%A.~Antognini, F.~Kottmann, F.~Biraben, P.~Indelicato, F.~Nez and R.~Pohl, \textit{{Theory of the 2S-2P Lamb shift and 2S hyperfine splitting in muonic hydrogen}}, Annals Phys. \textbf{331}, 127 (2013), \doi{10.1016/j.aop.2012.12.003}, \eprint{1208.2637}.

A. Antognini, F. Kottmann, F. Biraben, P. Indelicato, F. Nez and R. Pohl, \textit{Theory of the 2S–2P Lamb shift and 2S hyperfine splitting in muonic hydrogen}, Ann. Phys. \textbf{331}, 127 (2013), \doi{10.1016/j.aop.2012.12.003}.




\bibitem{Alarcon:2020kcz} 
%\textcolor{red}{TODO}
%J.~M. Alarc\'on, D.~W. Higinbotham and C.~Weiss, \textit{{Precise determination of the proton magnetic radius from electron scattering data}}, Phys. Rev. C \textbf{102}(3), 035203 (2020), \doi{10.1103/PhysRevC.102.035203}, \eprint{2002.05167}.

J. M. Alarcón, D. W. Higinbotham and C. Weiss, \textit{Precise determination of the proton magnetic radius from electron scattering data}, Phys. Rev. C \textbf{102}, 035203 (2020), \doi{10.1103/PhysRevC.102.035203}.




\bibitem{Maximon_2000} 
%\textcolor{red}{TODO}
%L.~C. Maximon and J.~A. Tjon, \textit{Radiative corrections to electron-proton scattering}, Physical Review C \textbf{62}(5) (2000), \doi{10.1103/physrevc.62.054320}.

L. C. Maximon and J. A. Tjon, \textit{Radiative corrections to electron-proton scattering}, Phys. Rev. C \textbf{62}, 054320 (2000), \doi{10.1103/physrevc.62.054320}.




\bibitem{Gramolin:2014pva} 
%\textcolor{red}{TODO}
%A.~Gramolin, V.~Fadin, A.~Feldman, R.~Gerasimov, D.~Nikolenko, I.~Rachek and D.~Toporkov, \textit{{A new event generator for the elastic scattering of charged leptons on protons}}, J. Phys. G \textbf{41}(11), 115001 (2014), \doi{10.1088/0954-3899/41/11/115001}, \eprint{1401.2959}.

A. V. Gramolin, V. S. Fadin, A. L. Feldman, R. E. Gerasimov, D. M. Nikolenko, I. A. Rachek and D. K. Toporkov, \textit{A new event generator for the elastic scattering of charged leptons on protons}, J. Phys. G: Nucl. Part. Phys. \textbf{41}, 115001 (2014), \doi{10.1088/0954-3899/41/11/115001}.




\bibitem{Akushevich:2015toa}
 %\textcolor{red}{TODO}
%I.~Akushevich, H.~Gao, A.~Ilyichev and M.~Meziane, \textit{{Radiative corrections beyond the ultra relativistic limit in unpolarized ep elastic and M{\o}ller scatterings for the PRad Experiment at Jefferson Laboratory}}, Eur. Phys. J. A \textbf{51}(1), 1 (2015), \doi{10.1140/epja/i2015-15001-8}.

I. Akushevich, H. Gao, A. Ilyichev and M. Meziane, \textit{Radiative corrections beyond the ultra relativistic limit in unpolarized ep elastic and Møller scatterings for the PRad Experiment at Jefferson Laboratory}, Eur. Phys. J. A \textbf{51}, 1 (2015), \doi{10.1140/epja/i2015-15001-8}.




\bibitem{Bucoveanu:2018soy} 
%\textcolor{red}{TODO}
%R.-D. Bucoveanu and H.~Spiesberger, \textit{{Second-Order Leptonic Radiative Corrections for Lepton-Proton Scattering}}, Eur. Phys. J. A \textbf{55}(4), 57 (2019), \doi{10.1140/epja/i2019-12727-1}, \eprint{1811.04970}.

R.-D. Bucoveanu and H. Spiesberger, \textit{Second-order leptonic radiative corrections for lepton-proton scattering}, Eur. Phys. J. A \textbf{55}, 57 (2019), \doi{10.1140/epja/i2019-12727-1}.




\bibitem{Banerjee:2020rww} 
%\textcolor{red}{TODO}
%P.~Banerjee, T.~Engel, A.~Signer and Y.~Ulrich, \textit{{QED at NNLO with McMule}}, SciPost Phys. \textbf{9}, 027 (2020), \doi{10.21468/SciPostPhys.9.2.027}, \eprint{2007.01654}.

P. Banerjee, T. Engel, A. Signer and Y. Ulrich, \textit{QED at NNLO with McMule}, SciPost Phys. \textbf{9}, 027 (2020), \doi{10.21468/SciPostPhys.9.2.027}.




\bibitem{Carlson:2007sp} 
%\textcolor{red}{TODO}
%C.~E. Carlson and M.~Vanderhaeghen, \textit{{Two-Photon Physics in Hadronic Processes}}, Ann. Rev. Nucl. Part. Sci. \textbf{57}, 171 (2007), \doi{10.1146/annurev.nucl.57.090506.123116}, \eprint{hep-ph/0701272}.

C. E. Carlson and M. Vanderhaeghen, \textit{Two-photon physics in hadronic processes}, Annu. Rev. Nucl. Part. Sci. \textbf{57}, 171 (2007), \doi{10.1146/annurev.nucl.57.090506.123116}.




\bibitem{Arrington:2011dn} 
%\textcolor{red}{TODO}
%J.~Arrington, P.~Blunden and W.~Melnitchouk, \textit{{Review of two-photon exchange in electron scattering}}, Prog. Part. Nucl. Phys. \textbf{66}, 782 (2011), \doi{10.1016/j.ppnp.2011.07.003}, \eprint{1105.0951}.

J. Arrington, P. G. Blunden and W. Melnitchouk, \textit{Review of two-photon exchange in electron scattering}, Prog. Part. Nucl. Phys. \textbf{66}, 782 (2011), \doi{10.1016/j.ppnp.2011.07.003}.




\bibitem{Afanasev:2017gsk} 
%\textcolor{red}{TODO}
%A.~Afanasev, P.~Blunden, D.~Hasell and B.~Raue, \textit{{Two-photon exchange in elastic electron--proton scattering}}, Prog. Part. Nucl. Phys. \textbf{95}, 245 (2017), \doi{10.1016/j.ppnp.2017.03.004}, \eprint{1703.03874}.

A. Afanasev, P. G. Blunden, D. Hasell and B. A. Raue, \textit{Two-photon exchange in elastic electron–proton scattering}, Prog. Part. Nucl. Phys. \textbf{95}, 245 (2017), \doi{10.1016/j.ppnp.2017.03.004}.




\bibitem{Tomalak:2017shs} 
%\textcolor{red}{TODO}
%O.~Tomalak, B.~Pasquini and M.~Vanderhaeghen, \textit{{Two-photon exchange contribution to elastic $e^-$-proton scattering: Full dispersive treatment of $\pi$N states and comparison with data}}, Phys. Rev. D \textbf{96}(9), 096001 (2017), \doi{10.1103/PhysRevD.96.096001}, \eprint{1708.03303}.

O. Tomalak, B. Pasquini and M. Vanderhaeghen, \textit{Two-photon exchange contribution to elastic ${e}^{\ensuremath{-}}$-proton scattering: Full dispersive treatment of $\pi N$ states and comparison with data}, Phys. Rev. D \textbf{96}, 096001 (2017), \doi{10.1103/PhysRevD.96.096001}.




\bibitem{Hill:2017wgb} 
%\textcolor{red}{TODO}
%R.~J. Hill, P.~Kammel, W.~J. Marciano and A.~Sirlin, \textit{{Nucleon Axial Radius and Muonic Hydrogen \textemdash{} A New Analysis and Review}}, Rept. Prog. Phys. \textbf{81}(9), 096301 (2018), \doi{10.1088/1361-6633/aac190}, \eprint{1708.08462}.

R. J. Hill, P. Kammel, W. J. Marciano and A. Sirlin, \textit{Nucleon axial radius and muonic hydrogen — a new analysis and review}, Rep. Prog. Phys. \textbf{81}, 096301 (2018), \doi{10.1088/1361-6633/aac190}.




\bibitem{Galster:1971kv} 
%\textcolor{red}{TODO}
%S.~Galster, H.~Klein, J.~Moritz, K.~H. Schmidt, D.~Wegener and J.~Bleckwenn, \textit{{Elastic electron-deuteron scattering and the electric neutron form factor at four-momentum transfers 5fm$^{-2} < q^2 < 14$fm$^{-2}$}}, Nucl. Phys. B \textbf{32}, 221 (1971), \doi{10.1016/0550-3213(71)90068-X}.

S. Galster, H. Klein, J. Moritz, K. H. Schmidt, D. Wegener and J. Bleckwenn, \textit{Elastic electron-deuteron scattering and the electric neutron form factor at four-momentum transfers $5 fm^{-2} < q^2 < 14 fm^{-2}$}, Nucl. Phys. B \textbf{32}, 221 (1971), \doi{10.1016/0550-3213(71)90068-X}.




\bibitem{Krauth:2015nja} 
%\textcolor{red}{TODO}
%J.~J. Krauth, M.~Diepold, B.~Franke, A.~Antognini, F.~Kottmann and R.~Pohl, \textit{{Theory of the n=2 levels in muonic deuterium}}, Annals Phys. \textbf{366}, 168 (2016), \doi{10.1016/j.aop.2015.12.006}, \eprint{1506.01298}.

J. J. Krauth, M. Diepold, B. Franke, A. Antognini, F. Kottmann and R. Pohl, \textit{Theory of the n=2 levels in muonic deuterium}, Ann. Phys. \textbf{366}, 168 (2016), \doi{10.1016/j.aop.2015.12.006}.




\bibitem{Franke:2017tpc} 
%\textcolor{red}{TODO}
%B.~Franke, J.~J. Krauth, A.~Antognini, M.~Diepold, F.~Kottmann and R.~Pohl, \textit{{Theory of the n = 2 levels in muonic helium-3 ions}}, Eur. Phys. J. D \textbf{71}(12), 341 (2017), \doi{10.1140/epjd/e2017-80296-1}, \eprint{1705.00352}.

B. Franke, J. J. Krauth, A. Antognini, M. Diepold, F. Kottmann and R. Pohl, \textit{Theory of the n = 2 levels in muonic helium-3 ions}, Eur. Phys. J. D \textbf{71}, 341 (2017), \doi{10.1140/epjd/e2017-80296-1}.




\bibitem{Diepold:2016cxv}
 %\textcolor{red}{TODO}
%%M.~Diepold, B.~Franke, J.~J. Krauth, A.~Antognini, F.~Kottmann and R.~Pohl, \textit{{Theory of the Lamb shift and Fine Structure in muonic $\mathrm{^4He}$ ions and the muonic $\mathrm{^3He-^4He}$ Isotope Shift}}, Annals Phys. \textbf{396}, 220 (2018), \doi{10.1016/j.aop.2018.07.015}, \eprint{1606.05231}.

M. Diepold, B. Franke, J. J. Krauth, A. Antognini, F. Kottmann and R. Pohl, \textit{Theory of the Lamb Shift and fine structure in muonic $^4$He ions and the muonic $^3$He–$^4$He Isotope Shift}, Ann. Phys. \textbf{396}, 220 (2018), \doi{10.1016/j.aop.2018.07.015}.




\bibitem{Zyla:2020zbs}
 %\textcolor{red}{TODO}

P. A. Zyla \textit{et al.}, \textit{{Review of particle physics}}, Prog. Theor. Exp. Phys., 083C01 (2020), \doi{10.1093/ptep/ptaa104}.

%None, \textit{Review of Particle Physics}, None \textbf{2020}, None (2020), \doi{10.1093/ptep/ptaa104}.




\bibitem{Phillips:1999hh} 
%\textcolor{red}{TODO}
%D.~R. Phillips, G.~Rupak and M.~J. Savage, \textit{{Improving the convergence of N N effective field theory}}, Phys. Lett. B \textbf{473}, 209 (2000), \doi{10.1016/S0370-2693(99)01496-3}, \eprint{nucl-th/9908054}.

D. R. Phillips, G. Rupak and M. J. Savage, \textit{Improving the convergence of NN effective field theory}, Phys. Lett. B \textbf{473}, 209 (2000), \doi{10.1016/S0370-2693(99)01496-3}.




\bibitem{Carlson:2013xea} 
%\textcolor{red}{TODO}
%C.~E. Carlson, M.~Gorchtein and M.~Vanderhaeghen, \textit{{Nuclear structure contribution to the Lamb shift in muonic deuterium}}, Phys. Rev. A \textbf{89}(2), 022504 (2014), \doi{10.1103/PhysRevA.89.022504}, \eprint{1311.6512}.

C. E. Carlson, M. Gorchtein and M. Vanderhaeghen, \textit{Nuclear-structure contribution to the Lamb shift in muonic deuterium}, Phys. Rev. A \textbf{89}, 022504 (2014), \doi{10.1103/PhysRevA.89.022504}.




\bibitem{Carlson:2016cii} 
%\textcolor{red}{TODO}
%C.~E. Carlson, M.~Gorchtein and M.~Vanderhaeghen, \textit{{Two-photon exchange correction to $2S-2P$ splitting in muonic $^3$He ions}}, Phys. Rev. A \textbf{95}(1), 012506 (2017), \doi{10.1103/PhysRevA.95.012506}, \eprint{1611.06192}.

C. E. Carlson, M. Gorchtein and M. Vanderhaeghen, \textit{Two-photon exchange correction to $2S-2P$ splitting in muonic $^3$He ions}, Phys. Rev. A \textbf{95}, 012506 (2017), \doi{10.1103/PhysRevA.95.012506}.




\bibitem{Pachucki:2015uga}
 %\textcolor{red}{TODO}
%K.~Pachucki and A.~Wienczek, \textit{{Nuclear structure effects in light muonic atoms}}, Phys. Rev. A \textbf{91}(4), 040503 (2015), \doi{10.1103/PhysRevA.91.040503}, \eprint{1501.07451}.

K. Pachucki and A. Wienczek, \textit{Nuclear structure effects in light muonic atoms}, Phys. Rev. A \textbf{91}, 040503 (2015), \doi{10.1103/PhysRevA.91.040503}.




\bibitem{Wiringa:1994wb}
 %\textcolor{red}{TODO}
%R.~B. Wiringa, V.~G.~J. Stoks and R.~Schiavilla, \textit{{An Accurate nucleon-nucleon potential with charge independence breaking}}, Phys. Rev. C \textbf{51}, 38 (1995), \doi{10.1103/PhysRevC.51.38}, \eprint{nucl-th/9408016}.

R. B. Wiringa, V. G. J. Stoks and R. Schiavilla, \textit{Accurate nucleon-nucleon potential with charge-independence breaking}, Phys. Rev. C \textbf{51}, 38 (1995), \doi{10.1103/PhysRevC.51.38}.




\bibitem{Ji:2013oba} 
%\textcolor{red}{TODO}
%C.~Ji, N.~Nevo~Dinur, S.~Bacca and N.~Barnea, \textit{{Nuclear Polarization Corrections to the $\mu^4$He$^+$ Lamb Shift}}, Phys. Rev. Lett. \textbf{111}, 143402 (2013), \doi{10.1103/PhysRevLett.111.143402}, \eprint{1307.6577}.

C. Ji, N. Nevo Dinur, S. Bacca and N. Barnea, \textit{Nuclear polarization corrections to the $\mu^4$He$^+$ Lamb Shift}, Phys. Rev. Lett. \textbf{111}, 143402 (2013), \doi{10.1103/PhysRevLett.111.143402}.




\bibitem{Hernandez:2014pwa}
 %\textcolor{red}{TODO}
%O.~J. Hernandez, C.~Ji, S.~Bacca, N.~Nevo~Dinur and N.~Barnea, \textit{{Improved estimates of the nuclear structure corrections in $\mu$D}}, Phys. Lett. B \textbf{736}, 344 (2014), \doi{10.1016/j.physletb.2014.07.039}, \eprint{1406.5230}.

O. J. Hernandez, C. Ji, S. Bacca, N. Nevo Dinur and N. Barnea, \textit{Improved estimates of the nuclear structure corrections in $\mu$D}, Phys. Lett. B \textbf{736}, 344 (2014), \doi{10.1016/j.physletb.2014.07.039}.




\bibitem{Dinur:2015vzv} 
%\textcolor{red}{TODO}
%N.~Nevo~Dinur, C.~Ji, S.~Bacca and N.~Barnea, \textit{{Nuclear structure corrections to the Lamb shift in $\mu^3$He$^+$ and $\mu^3$H}}, Phys. Lett. B \textbf{755}, 380 (2016), \doi{10.1016/j.physletb.2016.02.023}, \eprint{1512.05773}.

N. Nevo Dinur, C. Ji, S. Bacca and N. Barnea, \textit{Nuclear structure corrections to the Lamb shift in $\mu^3$He$^+$ and $\mu^3$H}, Phys. Lett. B \textbf{755}, 380 (2016), \doi{10.1016/j.physletb.2016.02.023}.




\bibitem{Ji:2018ozm} 
%\textcolor{red}{TODO}
%C.~Ji, S.~Bacca, N.~Barnea, O.~J. Hernandez and N.~Nevo-Dinur, \textit{{$Ab initio$ calculation of nuclear structure corrections in muonic atoms}}, J. Phys. G \textbf{45}(9), 093002 (2018), \doi{10.1088/1361-6471/aad3eb}, \eprint{1806.03101}.

C. Ji, S. Bacca, N. Barnea, O. J. Hernandez and N. Nevo Dinur, \textit{Ab initio calculation of nuclear-structure corrections in muonic atoms}, J. Phys. G: Nucl. Part. Phys. \textbf{45}, 093002 (2018), \doi{10.1088/1361-6471/aad3eb}.




\bibitem{Pastore:2014iga} 
%\textcolor{red}{TODO}
%S.~Pastore, F.~Myhrer and K.~Kubodera, \textit{{An update of muon capture on hydrogen}}, Int. J. Mod. Phys. E \textbf{23}(08), 1430010 (2014), \doi{10.1142/S0218301314300100}, \eprint{1405.1358}.

S. Pastore, F. Myhrer and K. Kubodera, \textit{An update of muon capture on hydrogen}, Int. J. Mod. Phys. E \textbf{23}, 1430010 (2014), \doi{10.1142/S0218301314300100}.




\bibitem{section04} 
%\textcolor{red}{TODO}

B. Lauss and B. Blau, \textit{{UCN, the ultracold neutron source -- neutrons for particle physics}}, SciPost Phys. Proc. \textbf{5}, 004 (2021), \doi{10.21468/SciPostPhysProc.5.004}.

%AUTHORS, \textit{TITLE}, JOURNAL \textbf{VOLUME}, PAGE/ARTICLE NUMBER (YEAR), \doi{DOI}.




\bibitem{Fornal:2018eol} 
%\textcolor{red}{TODO}
%B.~Fornal and B.~Grinstein, \textit{{Dark Matter Interpretation of the Neutron Decay Anomaly}}, Phys. Rev. Lett. \textbf{120}(19), 191801 (2018), \doi{10.1103/PhysRevLett.120.191801}, [Erratum: Phys.Rev.Lett. 124, 219901 (2020)], \eprint{1801.01124}.

B. Fornal and B. Grinstein, \textit{Dark matter interpretation of the neutron decay anomaly}, Phys. Rev. Lett. \textbf{120}, 191801 (2018), \doi{10.1103/PhysRevLett.120.191801}.




\bibitem{Czarnecki:2018okw} 
%\textcolor{red}{TODO}
%A.~Czarnecki, W.~J. Marciano and A.~Sirlin, \textit{{Neutron Lifetime and Axial Coupling Connection}}, Phys. Rev. Lett. \textbf{120}(20), 202002 (2018), \doi{10.1103/PhysRevLett.120.202002}, \eprint{1802.01804}.

A. Czarnecki, W. J. Marciano and A. Sirlin, \textit{Neutron lifetime and axial coupling connection}, Phys. Rev. Lett. \textbf{120}, 202002 (2018), \doi{10.1103/PhysRevLett.120.202002}.




\bibitem{Lee:1956qn} 
%\textcolor{red}{TODO}
%T.~D. Lee and C.-N. Yang, \textit{{Question of Parity Conservation in Weak Interactions}}, Phys. Rev. \textbf{104}, 254 (1956), \doi{10.1103/PhysRev.104.254}.

T. D. Lee and C. N. Yang, \textit{Question of parity conservation in weak interactions}, Phys. Rev. \textbf{104}, 254 (1956), \doi{10.1103/PhysRev.104.254}.




\bibitem{Severijns:2006dr} 
%\textcolor{red}{TODO}
%N.~Severijns, M.~Beck and O.~Naviliat-Cuncic, \textit{{Tests of the standard electroweak model in beta decay}}, Rev. Mod. Phys. \textbf{78}, 991 (2006), \doi{10.1103/RevModPhys.78.991}, \eprint{nucl-ex/0605029}.

N. Severijns, M. Beck and O. Naviliat-Cuncic, \textit{Tests of the standard electroweak model in nuclear beta decay}, Rev. Mod. Phys. \textbf{78}, 991 (2006), \doi{10.1103/RevModPhys.78.991}.




\bibitem{Hagelstein:2015egb} 
%\textcolor{red}{TODO}
%F.~Hagelstein, R.~Miskimen and V.~Pascalutsa, \textit{{Nucleon Polarizabilities: from Compton Scattering to Hydrogen Atom}}, Prog. Part. Nucl. Phys. \textbf{88}, 29 (2016), \doi{10.1016/j.ppnp.2015.12.001}, \eprint{1512.03765}.

F. Hagelstein, R. Miskimen and V. Pascalutsa, \textit{Nucleon polarizabilities: From Compton scattering to hydrogen atom}, Prog. Part. Nucl. Phys. \textbf{88}, 29 (2016), \doi{10.1016/j.ppnp.2015.12.001}.




\bibitem{Ananthanarayan:2000ht} 
%\textcolor{red}{TODO}
%B.~Ananthanarayan, G.~Colangelo, J.~Gasser and H.~Leutwyler, \textit{{Roy equation analysis of pi pi scattering}}, Phys. Rept. \textbf{353}, 207 (2001), \doi{10.1016/S0370-1573(01)00009-6}, \eprint{hep-ph/0005297}.

B. Ananthanarayan, G. Colangelo, J. Gasser and H. Leutwyler, \textit{Roy equation analysis of $\pi\pi$ scattering}, Phys. Rep. \textbf{353}, 207 (2001), \doi{10.1016/S0370-1573(01)00009-6}.




\bibitem{Caprini:2011ky} 
%\textcolor{red}{TODO}
%I.~Caprini, G.~Colangelo and H.~Leutwyler, \textit{{Regge analysis of the pi pi scattering amplitude}}, Eur. Phys. J. C \textbf{72}, 1860 (2012), \doi{10.1140/epjc/s10052-012-1860-1}, \eprint{1111.7160}.

I. Caprini, G. Colangelo and H. Leutwyler, \textit{Regge analysis of the $\pi\pi$ scattering amplitude}, Eur. Phys. J. C \textbf{72}, 1860 (2012), \doi{10.1140/epjc/s10052-012-1860-1}.




\bibitem{GarciaMartin:2011cn} 
%\textcolor{red}{TODO}
%R.~Garcia-Martin, R.~Kaminski, J.~R. Pelaez, J.~Ruiz~de Elvira and F.~J. Yndurain, \textit{{The Pion-pion scattering amplitude. IV: Improved analysis with once subtracted Roy-like equations up to 1100 MeV}}, Phys. Rev. D \textbf{83}, 074004 (2011), \doi{10.1103/PhysRevD.83.074004}, \eprint{1102.2183}.

R. García-Martín, R. Kamiński, J. R. Peláez, J. Ruiz de Elvira and F. J. Ynduráin, \textit{Pion-pion scattering amplitude. IV. Improved analysis with once subtracted Roy-like equations up to 1100 MeV}, Phys. Rev. D \textbf{83}, 074004 (2011), \doi{10.1103/PhysRevD.83.074004}.




\bibitem{Gasser:2007zt} 
%\textcolor{red}{TODO}
%J.~Gasser, V.~E. Lyubovitskij and A.~Rusetsky, \textit{{Hadronic atoms in QCD + QED}}, Phys. Rept. \textbf{456}, 167 (2008), \doi{10.1016/j.physrep.2007.09.006}, \eprint{0711.3522}.

J. Gasser, V. E. Lyubovitskij and A. Rusetsky, \textit{Hadronic atoms in QCD+QED}, Phys. Rep. \textbf{456}, 167 (2008), \doi{10.1016/j.physrep.2007.09.006}.




\bibitem{Hoferichter:2015hva} 
%\textcolor{red}{TODO}
%M.~Hoferichter, J.~Ruiz~de Elvira, B.~Kubis and U.-G. Mei\ss{}ner, \textit{{Roy\textendash{}Steiner-equation analysis of pion\textendash{}nucleon scattering}}, Phys. Rept. \textbf{625}, 1 (2016), \doi{10.1016/j.physrep.2016.02.002}, \eprint{1510.06039}.

M. Hoferichter, J. Ruiz de Elvira, B. Kubis and U.-G. Meißner, \textit{Roy–Steiner-equation analysis of pion–nucleon scattering}, Phys. Rep. \textbf{625}, 1 (2016), \doi{10.1016/j.physrep.2016.02.002}.




\bibitem{Das:1967it} 
%\textcolor{red}{TODO}
%T.~Das, G.~S. Guralnik, V.~S. Mathur, F.~E. Low and J.~E. Young, \textit{{Electromagnetic mass difference of pions}}, Phys. Rev. Lett. \textbf{18}, 759 (1967), \doi{10.1103/PhysRevLett.18.759}.

T. Das, G. S. Guralnik, V. S. Mathur, F. E. Low and J. E. Young, \textit{Electromagnetic mass difference of pions}, Phys. Rev. Lett. \textbf{18}, 759 (1967), \doi{10.1103/PhysRevLett.18.759}.




\bibitem{Donoghue:1996zn} 
%\textcolor{red}{TODO}
%J.~F. Donoghue and A.~F. Perez, \textit{{The Electromagnetic mass differences of pions and kaons}}, Phys. Rev. D \textbf{55}, 7075 (1997), \doi{10.1103/PhysRevD.55.7075}, \eprint{hep-ph/9611331}.

J. F. Donoghue and A. F. Pérez, \textit{Electromagnetic mass differences of pions and kaons}, Phys. Rev. D \textbf{55}, 7075 (1997), \doi{10.1103/PhysRevD.55.7075}.




\bibitem{Ananthanarayan:2002kj} 
%\textcolor{red}{TODO}
%B.~Ananthanarayan and B.~Moussallam, \textit{{Electromagnetic corrections in the anomaly sector}}, JHEP \textbf{05}, 052 (2002), \doi{10.1088/1126-6708/2002/05/052}, \eprint{hep-ph/0205232}.

B. Ananthanarayan and B. Moussallam, \textit{Electromagnetic corrections in the anomaly sector}, J. High Energy Phys. \textbf{05}, 052 (2002), \doi{10.1088/1126-6708/2002/05/052}.




\bibitem{Kampf:2009tk} 
%\textcolor{red}{TODO}
%K.~Kampf and B.~Moussallam, \textit{{Chiral expansions of the pi0 lifetime}}, Phys. Rev. D \textbf{79}, 076005 (2009), \doi{10.1103/PhysRevD.79.076005}, \eprint{0901.4688}.

K. Kampf and B. Moussallam, \textit{Chiral expansions of the $\pi^0$ lifetime}, Phys. Rev. D \textbf{79}, 076005 (2009), \doi{10.1103/PhysRevD.79.076005}.




\bibitem{Masjuan:2017tvw} 
%\textcolor{red}{TODO}
%P.~Masjuan and P.~Sanchez-Puertas, \textit{{Pseudoscalar-pole contribution to the $(g_{\mu}-2)$: a rational approach}}, Phys. Rev. D \textbf{95}(5), 054026 (2017), \doi{10.1103/PhysRevD.95.054026}, \eprint{1701.05829}.

P. Masjuan and P. Sanchez-Puertas, \textit{Pseudoscalar-pole contribution to the $(g_{\mu}-2)$: A rational approach}, Phys. Rev. D \textbf{95}, 054026 (2017), \doi{10.1103/PhysRevD.95.054026}.




\bibitem{Hoferichter:2018kwz} 
%\textcolor{red}{TODO}
%M.~Hoferichter, B.-L. Hoid, B.~Kubis, S.~Leupold and S.~P. Schneider, \textit{{Dispersion relation for hadronic light-by-light scattering: pion pole}}, JHEP \textbf{10}, 141 (2018), \doi{10.1007/JHEP10(2018)141}, \eprint{1808.04823}.

M. Hoferichter, B.-L. Hoid, B. Kubis, S. Leupold and S. P. Schneider, \textit{Dispersion relation for hadronic light-by-light scattering: pion pole}, J. High Energy Phys. \textbf{10}, 141 (2018), \doi{10.1007/JHEP10(2018)141}.




\bibitem{Gerardin:2019vio} 
%\textcolor{red}{TODO}
%A.~G\'erardin, H.~B. Meyer and A.~Nyffeler, \textit{{Lattice calculation of the pion transition form factor with $N_f=2+1$ Wilson quarks}}, Phys. Rev. D \textbf{100}(3), 034520 (2019), \doi{10.1103/PhysRevD.100.034520}, \eprint{1903.09471}.

A. Gérardin, H. B. Meyer and A. Nyffeler, \textit{Lattice calculation of the pion transition form factor with $N_f=2+1$ Wilson quarks}, Phys. Rev. D \textbf{100}, 034520 (2019), \doi{10.1103/PhysRevD.100.034520}.




\bibitem{Bryman:1982et} 
%\textcolor{red}{TODO}
%D.~A. Bryman, P.~Depommier and C.~Leroy, \textit{{PI ---\ensuremath{>} E neutrino, PI ---\ensuremath{>} E neutrino gamma decays and related processes}}, Phys. Rept. \textbf{88}, 151 (1982), \doi{10.1016/0370-1573(82)90162-4}.

D. A. Bryman, P. Depommier and C. Leroy, \textit{$\pi\rightarrow e\nu$, $\pi\rightarrow e\nu \gamma$ decays and related processes}, Phys. Rep. \textbf{88}, 151 (1982), \doi{10.1016/0370-1573(82)90162-4}.




\bibitem{Cirigliano:2002ng}
 %\textcolor{red}{TODO}
%V.~Cirigliano, M.~Knecht, H.~Neufeld and H.~Pichl, \textit{{The Pionic beta decay in chiral perturbation theory}}, Eur. Phys. J. C \textbf{27}, 255 (2003), \doi{10.1140/epjc/s2002-01093-2}, \eprint{hep-ph/0209226}.

V. Cirigliano, M. Knecht, H. Neufeld and H. Pichl, \textit{The pionic beta decay in chiral perturbation theory}, Eur. Phys. J. C \textbf{27}, 255 (2003), \doi{10.1140/epjc/s2002-01093-2}.




\bibitem{Cirigliano:2013xha} 
%\textcolor{red}{TODO}
%V.~Cirigliano, S.~Gardner and B.~Holstein, \textit{{Beta Decays and Non-Standard Interactions in the LHC Era}}, Prog. Part. Nucl. Phys. \textbf{71}, 93 (2013), \doi{10.1016/j.ppnp.2013.03.005}, \eprint{1303.6953}. %%%%%%%%%% END TODO: BBL

V. Cirigliano, S. Gardner and B. R. Holstein, \textit{Beta decays and non-standard interactions in the LHC era}, Prog. Part. Nucl. Phys. \textbf{71}, 93 (2013), \doi{10.1016/j.ppnp.2013.03.005}.



\end{thebibliography}

\end{document}